\documentclass[twocolumn]{aastex62}
\newcommand{\Mgas}{\ensuremath{M_{\rm{H_2}}}}
\newcommand{\fgas}{\ensuremath{f_{\rm{H_2}}}}
\newcommand{\tdep}{\ensuremath{t_{\rm{dep}}}}

\newcommand\squiggle{SQuIGG$\vec{L}$E\,\,}

\newcommand{\etal}{et al.}

\def\gtrsim{\mathrel{\hbox{\rlap{\hbox{\lower4pt\hbox{$\sim$}}}\hbox{\raise2pt\hbox{$>$}}}}}

\newcommand{\halpha}{H\ensuremath{\alpha}}

\newcommand{\hbeta}{H\ensuremath{\beta}}

\newcommand{\kms}{km~s\ensuremath{^{-1}}}

\newcommand{\msun}{\ensuremath{M_{\odot}}}
\newcommand{\mstar}{\ensuremath{M_*}}

\newcommand{\oii}{[\ion{O}{2}]}
\newcommand{\oiii}{[\ion{O}{3}]}

\graphicspath{{./}{figures/}}

\submitjournal{ApJ}

\shorttitle{Now you see it, now you don't: H$_2$ in massive post-starburst galaxies at $z\sim0.6$}
\shortauthors{Bezanson et al.}

\begin{document}

\title{Now you see it, now you don't: Star formation truncation precedes the loss of molecular gas by $\sim$100 Myr in massive post-starburst galaxies at z$\sim$0.6}

\author[0000-0001-5063-8254]{Rachel Bezanson}
\affiliation{Department of Physics and Astronomy and PITT PACC, University of Pittsburgh, Pittsburgh, PA, 15260, USA}

\author[0000-0003-3256-5615]{Justin S. Spilker}
\altaffiliation{NHFP Hubble Fellow}
\affiliation{Department of Astronomy, University of Texas at Austin, 2515 Speedway, Stop C1400, Austin, TX 78712, USA}
\affiliation{Department of Physics and Astronomy and George P. and Cynthia Woods Mitchell Institute for Fundamental Physics and Astronomy, Texas A\&M University, 4242 TAMU, College Station, TX 77843-4242}

\author[0000-0002-1714-1905]{Katherine A. Suess}
\affiliation{Astronomy Department, University of California, Berkeley, CA 94720, USA} 
\affiliation{Department of Astronomy and Astrophysics, University of California, Santa Cruz, 1156 High Street, Santa Cruz, CA 95064 USA}
\affiliation{Kavli Institute for Particle Astrophysics and Cosmology and Department of Physics, Stanford University, Stanford, CA 94305, USA}

\author[0000-0003-4075-7393]{David J. Setton}
\affiliation{Department of Physics and Astronomy and PITT PACC, University of Pittsburgh, Pittsburgh, PA, 15260, USA} 

\author[0000-0002-1109-1919]{Robert Feldmann}
\affiliation{Institute for Computational Science, University of Zurich, CH-8057 Zurich, Switzerland}

\author[0000-0002-5612-3427]{Jenny E. Greene}
\affiliation{Department of Astrophysical Sciences, Princeton University, Princeton, NJ 08544, USA}

\author[0000-0002-7613-9872]{Mariska Kriek} 
\affiliation{Leiden Observatory, Leiden University, P.O.Box 9513, NL-2300 AA Leiden, The Netherlands }
\affiliation{Astronomy Department, University of California, Berkeley, CA 94720, USA}

\author[0000-0002-7064-4309]{Desika Narayanan}
\affiliation{Department of Astronomy, University of Florida, 211 Bryant Space Science Center, Gainesville, FL, 32611, USA}
\affiliation{University of Florida Informatics Institute, 432 Newell Drive, CISE Bldg E251 Gainesville, FL, 32611, US}
\affiliation{Cosmic Dawn Centre at the Niels Bohr Institue, University of Copenhagen and DTU-Space, Technical University of Denmark}

\author{Margaret Verrico}
\affiliation{Department of Physics and Astronomy and PITT PACC, University of Pittsburgh, Pittsburgh, PA, 15260, USA} 

\correspondingauthor{Rachel Bezanson}
\email{rachel.bezanson@pitt.edu}

\begin{abstract}

We use ALMA observations of CO(2--1) in 13 massive ($\mstar\gtrsim 10^{11} \msun$) post-starburst galaxies at $z\sim0.6$ to constrain the molecular gas content in galaxies shortly after they quench their major star-forming episode. The post-starburst galaxies in this study are selected from the Sloan Digital Sky Survey spectroscopic samples (DR14) based on their spectral shapes, as part of the \squiggle program. Early results showed that two post-starburst galaxies host large $H_2$ reservoirs despite their low inferred star formation rates. Here we expand this analysis to a larger statistical sample of 13 galaxies. Six of the primary targets (45\%) are detected, with $\Mgas\gtrsim10^9$\msun. Given their high stellar masses, this mass limit corresponds to an average gas fraction of $\langle\fgas \equiv\Mgas/\mstar \rangle \sim7\%$, or ${\sim}14\%$ using lower stellar masses estimates derived from analytic, exponentially declining star formation histories. The gas fraction correlates with the $D_n4000$ spectral index, suggesting that the cold gas reservoirs decrease with time since burst, as found in local K+A galaxies. Star formation histories derived from flexible stellar population synthesis modeling support this empirical finding: galaxies that quenched $\lesssim 150$ Myr prior to observation host detectable CO(2-1) emission, while older post-starburst galaxies are undetected. The large $\mathrm{H_2}$ reservoirs and low {  star formation rates} in the sample imply that the quenching of star formation \emph{precedes} the disappearance of the cold gas reservoirs. However, within the following 100-200 Myrs, the \squiggle galaxies require the  additional and efficient heating or removal of cold gas to bring their low  { star formation rates} in line with standard  $\mathrm{H_2}$ scaling relations. 
\end{abstract}


\section{Introduction} \label{sec:intro}

The process by which star-forming massive disk galaxies shut off their on-going star formation and join the older population of quiescent elliptical galaxies is one of the most poorly understood aspects of galaxy evolution. The dominance of ellipticals and the relative inefficiency of star formation as measured by abundance-matching of the dark matter halo and galaxy mass functions above $M^{\star}$ has led to the need to introduce additional ``feedback'' into the galaxy formation process.  At the massive end, this feedback is generally attributed to active galactic nuclei (AGN) \citep[e.g.,][]{croton:06}. Most modern cosmological simulations that form realistic populations of massive galaxies introduce some mode of energy-injection that is attributed to supermassive black holes \citep[e.g.,][]{Crain:15,schaye:15,weinberger:17,anglesalcazar:17, pillepich:18, dave:19, rodriguezmontero:19}, but in other cases, the inefficiency to accrete and replenish the cold gas supplies of the most massive systems is tied to the halo mass \citep[e.g.,][]{feldmann:15,dave:17, feldmann:17}. In all cases, the link between cold $H_2$ and on-going star formation is a built-in assumption. These expectations are empirically well-motivated at large scales by the relatively gas-rich nature of galaxies with on-going star formation \citep[e.g.,][]{saintonge:11,saintonge:11b,saintonge:12,tacconi:13,tacconi:18} and at small scales by the strong correlation between the surface density of active star formation and the density of molecular hydrogen \citep[e.g,][]{kennicutt:98, schruba:11}. Although $H_2$ reservoirs in dynamically hot elliptical galaxies appears to be less efficient at fueling their low-level star formation, this effect is secondary; in the local Universe quiescent galaxies are extremely gas-depleted \citep[e.g.,][]{young:11,davis:11,davis:13}.

Empirical studies suggest that the majority of massive elliptical galaxies formed their stars in early, short-lived episodes - indicating the importance of a rapid mode of quenching \citep[e.g.][]{thomas:05,pacifici:16,tacchella:21}. It is therefore interesting to investigate the properties of post-starburst galaxies, sometimes referred to as E+A or K+A galaxies, which are selected to be the direct products of a fast-track of quenching that shut off a dramatic episode of star formation within $\lesssim1$ Gyr \citep{dresslergunn:83,zabludoff:96}. This truncation produces characteristic spectral signatures originating from A-stars\footnote{We note that strong Balmer absorption features are also apparent in BV and later-type stars, but adopt the convention of referring to these signatures as A-type signatures. This becomes important for some galaxies in the \squiggle survey for which inferred post-quenching ages are shorter than the lifetimes of A stars.} that dominate after more massive stars have died combined with a lack of signatures of instantaneous star formation (e.g., [OII] or H$\alpha$ emission lines). If indeed these galaxies are in transition and star formation has already been shut down a reasonable expectation would be that their gas reservoirs already resemble those of older quiescent galaxies. 

In this context, it is surprising that post-starburst galaxies have been demonstrated to host enigmatic and significant $\mathrm{H_2}$ reservoirs ($\fgas\equiv\Mgas/\mstar$ up to $\sim50\%$) even after their star formation stops abruptly. The majority of studies of $H_2$ in post-starburst galaxies have been limited to the local Universe, where such galaxies are extremely rare, but detailed studies are relatively accessible \citep[e.g.,][]{french:15, rowlands:15, alatalo:15, alatalo:16}. Given that the most massive galaxies have the oldest stellar populations and therefore quenched at the earliest epochs \citep[e.g.,][]{thomas:05,mcdermid:15}, these low redshift relics of the late-time quenching process do not necessarily reflect the processes that shut off star formation at early times.  Intriguingly, $H_2$ in local post-starburst galaxies has been demonstrated to disappear on short ($\sim$100 Myr) timescales \citep{french:18b}. However, no similar tests have been conducted beyond the local Universe. 

Pushing observations of the possible link -- or lack thereof -- between quenching and the availability of $H_2$ to an epoch when galaxies are shutting down their primary episodes of star formation is a critical test of the theoretical models that could form the most massive galaxies in the Universe. Only a handful of quiescent galaxies at $z>0.1$ have been targeted for $H_2$ using CO lines, yielding only four detections and/but deep non-detections of (older) quiescent galaxies at ``cosmic noon'' ($1\lesssim z\lesssim 2$) \citep{sargent:15, spilker:18, bezanson:19, belli:21, williams:21}. In a pilot study of two massive post-starburst galaxies at $z\sim0.6$ that were spectroscopically selected to have recently shut down a major episode of star formation, \citet{suess:17} reported both retained significant molecular gas reservoirs ($f_{H_2}\sim10-30\%$).

In this paper we expand upon that work with a larger sample of 13 galaxies from the \squiggle (Studying QUenching at Intermediate-z Galaxies: Gas, angu$\vec{L}$ar momentum, and Evolution) survey. This enables an investigation of trends between $H_2$ and the spectroscopic properties of massive post-starburst galaxies at $z\sim0.6$, immediately after they quench their { dominant} episode of star formation. In \S \ref{sec:data}, we briefly describe the \squiggle sample and the ALMA CO(2--1) observations. \S \ref{sec:gas} discusses the $H_2$ reservoirs implied by the CO(2--1) fluxes and puts the sample in the context of other studies of star-forming and post-starburst galaxies. Finally \S \ref{sec:discussion} provides a discussion of the results and avenues for future study. Throughout this paper we assume a concordance $\Lambda CDM$ cosmology with $\Omega_\Lambda=0.7$, $\Omega_m=0.3$ and $H_0=70$ \kms $\mathrm{Mpc^{-1}}$, a \citet{chabrier:03} initial mass function, and quote AB magnitudes.

\section{Data} \label{sec:data}

\subsection{The \squiggle Sample}

\begin{figure*}
\includegraphics[width=0.5\textwidth]{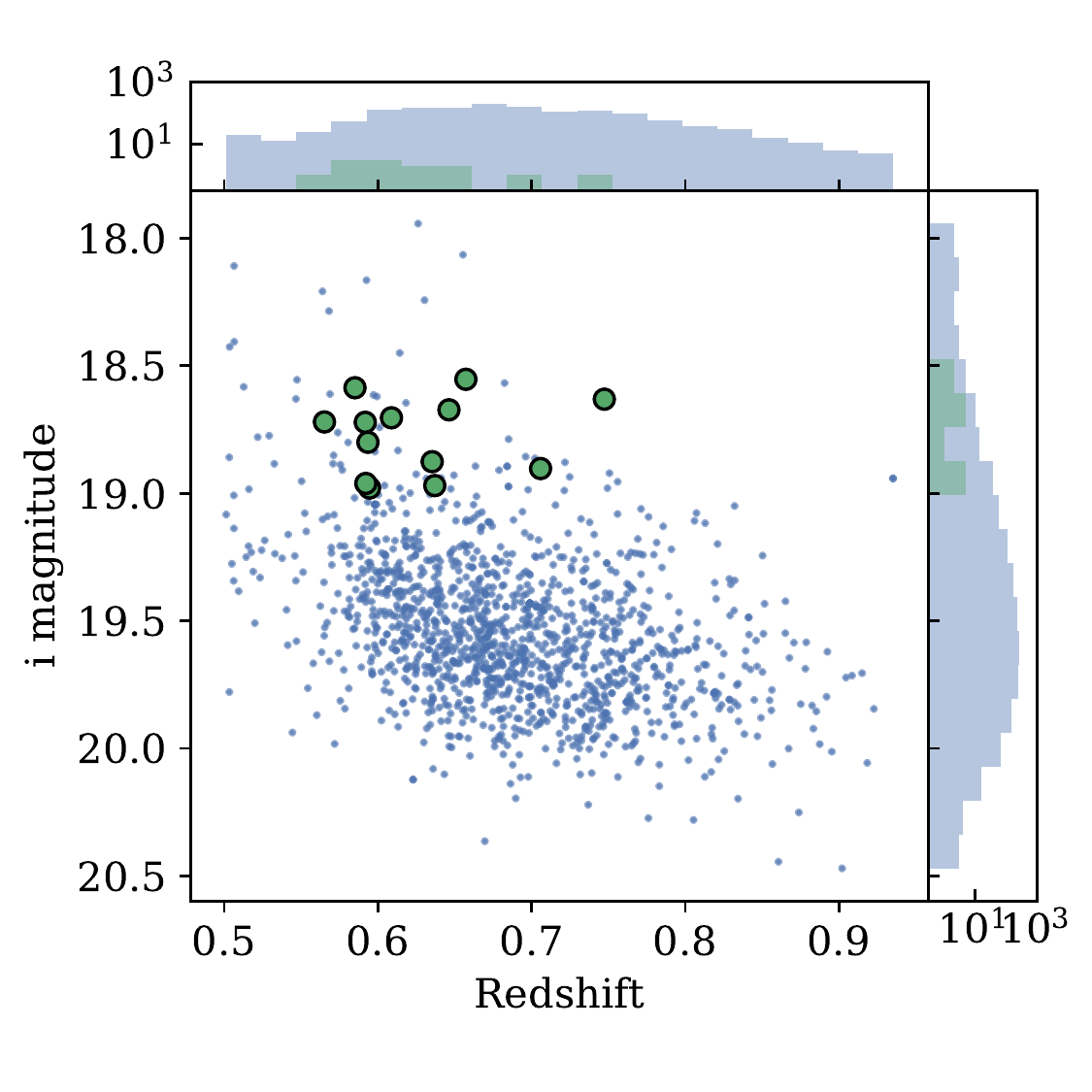}
\includegraphics[width=0.5\textwidth]{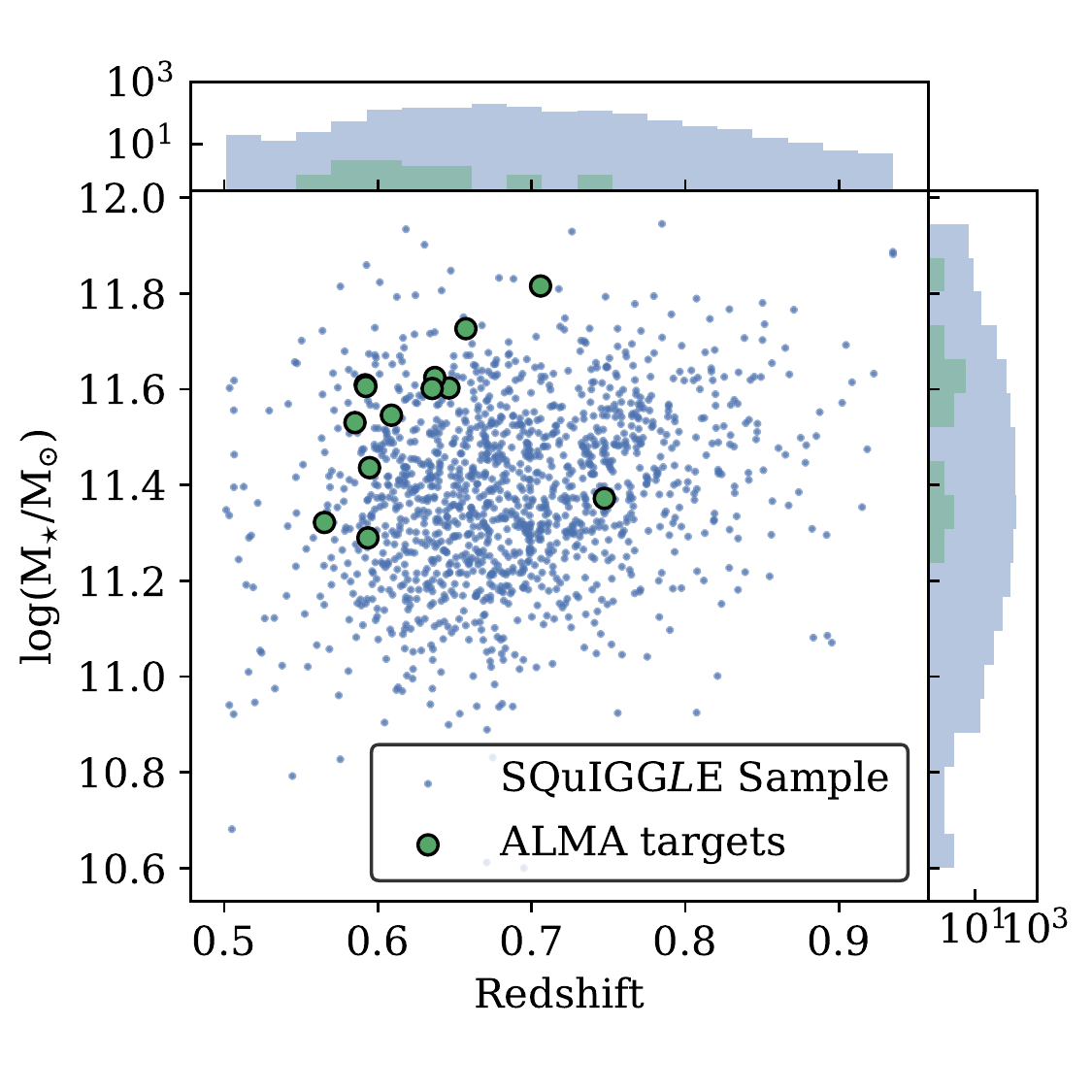}
    \caption{The distribution of \squiggle massive post-starburst galaxies in i magnitude (left) and stellar mass (right) versus spectroscopic redshift. In all panels the full sample is indicated by the blue symbols and histograms and the 13 galaxies targeted by our ALMA observations are highlighted in green. The ALMA targets are skewed towards slightly lower redshifts than the full sample ($z<0.75$) to catch the CO(2--1) line in ALMA band 4. ALMA targets were selected to have higher stellar masses and brighter magnitudes to allow for efficient multi-wavelength follow-up.
    \label{fig:distributions}}

\end{figure*}

The \squiggle sample is selected from the Sloan Digital Sky Survey DR14 spectroscopic database \citep{abolfathi:18} to have strong Balmer breaks and blue slopes redward of the break using rest-frame filters, following \citet{kriek:10}. For a detailed description of the spectroscopic identification and stellar populations of \squiggle galaxies, we refer the reader to the survey paper (K. Suess et al., submitted). In summary, for all objects in the database with $z>0.5$, each spectrum is integrated within three medium-width synthetic rest-frame filters ($U_m$, $B_m$, and $V_m$), which span the Balmer/4000$\mathrm{\AA}$ break and the spectral shape just redward of the break.  We only include galaxies with $S/N>6$ in the $B_m$ and $V_m$ fluxes and colors similar to those of A-type stars {($U_m-B_m > 0.975$ and $-0.25 < B_m-V_m < 0.45$)}. This selection yields a sample of 1318 unique objects, which span $0.50<z<0.94$. Figure \ref{fig:distributions} shows the distribution of the full \squiggle sample in i magnitude (left) and stellar mass (right) versus redshift. 

For this work, we rely on two sets of stellar population synthesis modeling, which we use in different contexts. The first set of models uses \texttt{Prospector} \citep{johnson:17, leja:17, johnson:21} to fit the SDSS $ugriz$ and WISE (3.4$\mu$m and $4.5\mu$m) photometry and spectra \citep{abolfathi:18, schlafly:19} with a custom set of ``non-parametric'' star formation histories { (SFHs)}, assuming a \citet{kriek:13} dust law. These star formation histories are similar to default continuity prior non-parametric models from \citet{leja:19}, with three fixed duration, variable { star formation rate (SFR)} bins prior to 2 Gyrs of lookback time, five equal-mass bins with variable edges that follow, and a unique final bin with flexibility in timing and SFR normalization. Without this final bin, the default continuity assumptions could be too strict to allow for dramatic bursts  or quick truncation in star formation as one might expect for post-starburst galaxies, effectively blurring out SFHs and biasing instantaneous SFRs to higher values. While conducting extensive recovery testing of stellar population properties using \texttt{Prospector}, we found that our adopted flexible star formation histories provide excellent recovery of instantaneous star formation rate (interpolated over the last 1 Myr) for sufficient SFR ($\mathrm{\gtrsim 1 M_{\odot}\, yr^{-1}}$), below which the measured star formation rates were poorly constrained by the existing spectra-photometric dataset (K. Suess et al. in prep). Therefore, when making comparisons to scaling relations, we set the star formation rates to a floor value of $1\mathrm{ M_{\odot}\,yr^{-1}}$ and label those points as upper limits. 

We note that the low SFRs are consistent with follow-up Keck/LRIS spectroscopy targeting \halpha\, (K. Suess et al., in prep). Although SFRs derived from \halpha\, luminosity are less uncertain than e.g. [OII] luminosity-based SFRs used in \citet{suess:17}, due to dust and other contaminating ionizing sources, H$\alpha$-based SFRs are insensitive to heavily dust-obscured star formation. We see no strong evidence for such extreme obscuration e.g., in the 2mm continuum data presented in this paper, but will return to this in \S \ref{sec:discussion}. 

This \texttt{Prospector} { spectral energy distribution (SED)} modeling is designed to accurately recover { SFHs} and { SFRs}, particularly immediately before quenching. However, often these histories are more extended then a more standard exponentially declining or delayed exponential analytic SFH, yielding significantly higher stellar masses. This $0.1-0.3$ dex offset is a generic consequence of ``non-parametric'' SED modeling \citep[see e.g.][]{leja:19,lower:20} and is perhaps more extreme for the post-starburst galaxies in this sample. While we expect that the higher stellar masses likely reflect the intrinsic properties of the galaxies, we also fit the SDSS spectra and photometry with delayed exponential star formation histories { \citep[as described in][]{setton:20}} assuming similar \citet{chabrier:03} IMF, \citet{bc:03} libraries, and a \citet{calzetti:97} dust law using \texttt{FAST++}, an implementation of the FAST {(Fitting and Assessment of Synthetic Templates)} software \citep{kriek:09}. The stellar masses derived from these fits ($M_{\star,FAST}$) are an average of 0.38 dex lower than the stellar masses derived in the default fits. We plan to expand upon these differences in an upcoming paper (K. Suess et al. in prep), but in this work adopt the lower values to place \squiggle\ galaxies on scaling relations for consistency. Regardless of the technique used to calculate stellar masses, galaxies in the \squiggle\, Survey are generally bright and massive ($\langle{\log \mstar/\msun}\rangle=11.4$), which is primarily driven by the spectroscopic signal-to-noise cut. 
\begin{figure}
    \centering
    \includegraphics[width=0.45\textwidth]{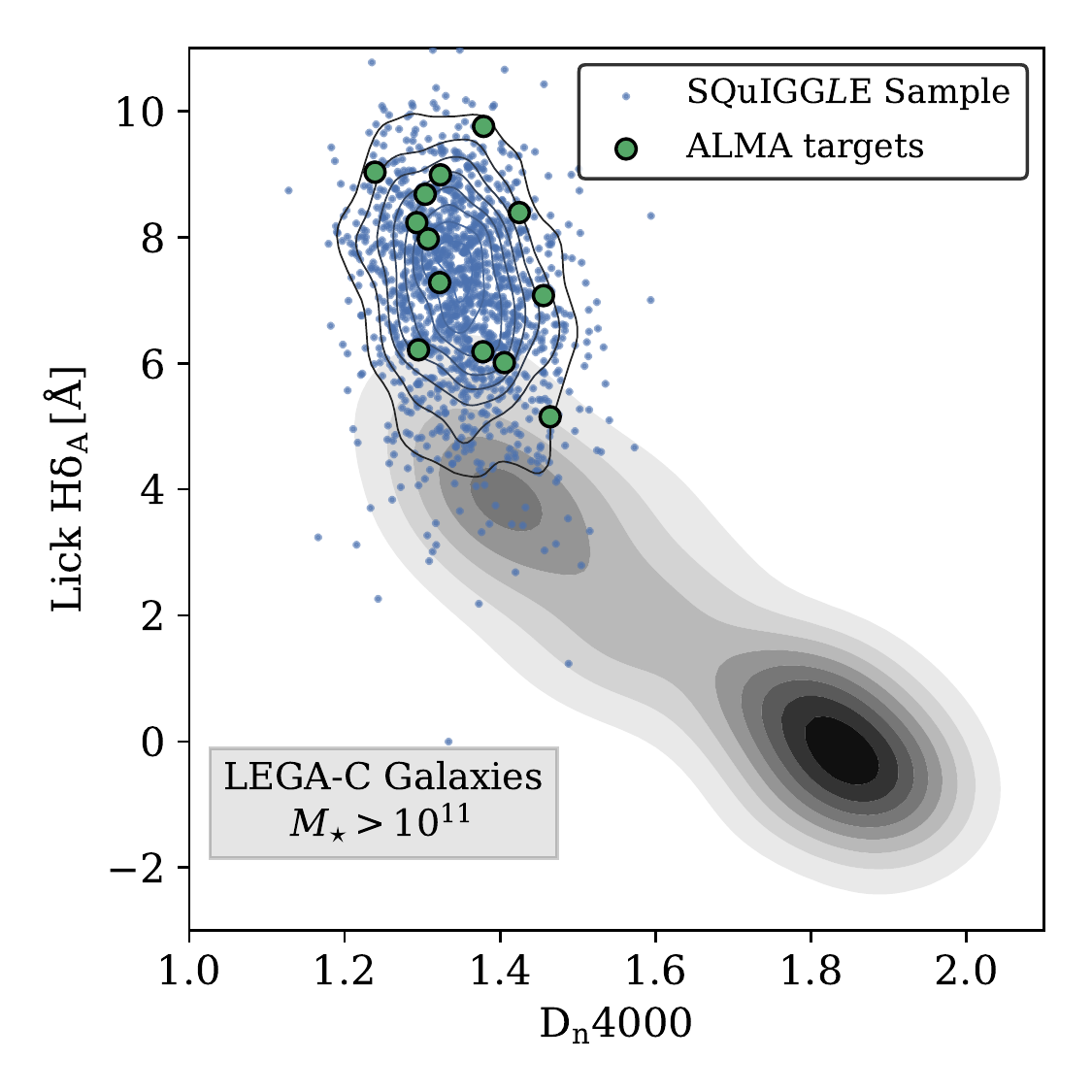}
    \caption{The distribution of the \squiggle sample in Lick $H\delta_A$ and $D_n4000$ spectral indices. Blue symbols indicate the full dataset and green circles highlight those targeted by our ALMA CO(2--1) survey. The distribution of a the population of similarly massive ($\log \mstar/\msun >11$) galaxies at $0.6<z<0.8$ from the LEGA-C survey are indicated by the grey shaded region. Although the selection of post-starburst galaxies in the \squiggle is based on their broad spectral shapes, the sample exhibits strong $H\delta_A$ absorption features and very blue $Dn4000$ spectral indices relative to the full population of massive galaxies at similar redshifts. The galaxies targeted in our ALMA survey span the range of H$\delta_A$ and D$_n4000$ parameter space spanned by the full \squiggle survey.
    \label{fig:hd_d4000_select}}
\end{figure}

Although these selection criteria are designed to identify galaxies immediately following the truncation of a significant episode of star formation based on the shape of their SEDs, it almost exclusively includes galaxies that would also be designated as post-starbursts based on strong H$\delta$ absorption (98\% of the sample has Lick H$\delta_A \geq 4.0\mathrm{\AA}$). For a comprehensive review of the range of post-starburst galaxy identification methods, we refer the reader to \citet{french:21}. Figure \ref{fig:hd_d4000_select} shows the \squiggle sample as small blue points in Lick H$\delta_A$ versus $D_n4000$ parameter space, which is commonly used to study the demographics of broad galaxy populations \citep[see e.g.,][]{kauffmann:03}.  { All indications from weak \oii\, emission and full spectral modeling indicate that galaxies in the \squiggle sample would pass cuts designed to identify objects without significant ongoing star formation, but the traditional $H\alpha$ SFR indicator is redshifted out of the spectral wavelength coverage for the full dataset.}  For comparison, we show the distribution of similarly massive ($\log M_{\star}/M_{\odot}>11$) galaxies at a similar epoch ($0.6<z<0.8$) from the third data release (DR3) of the Large Early Galaxy Astrophysics Census (LEGA-C) \citep{wel:16,straatman:17,wel:21} in gray contours. As demonstrated in \citet{wu:18}, massive galaxies at this epoch tend to have relatively old stellar populations as evidenced by weak Balmer absorption and strong 4000$\mathrm{\AA}$ breaks, corresponding to high $D_n4000$ spectral indices. This is in stark contrast with the recently-quenched post-starburst galaxies in the \squiggle survey; the LEGA-C galaxy distribution only barely reaches the low end of the \squiggle range in H$\delta_A$ indices. Although these samples probe similar redshifts, this lack of overlap is primarily an effect of survey volume. The $\sim$1.6 square degree section of the COSMOS field that was targeted by LEGA-C is relatively large for an extragalactic field, but it probes a vastly smaller volume than the BOSS Survey (10,000 square degrees) \citep{dawson:13}. 

\begin{deluxetable*}{ccccccc}[!t]
\tabletypesize{\scriptsize}
\tablecaption{Properties of the ALMA observations\label{tbl:alma}}
\tablehead{
    \colhead{ID} & \colhead{RA} & \colhead{Dec} & \colhead{ALMA program} & \colhead{Observation Date} & \colhead{Integration Time} & \colhead{Angular Resolution} \\
    \colhead{} & \colhead{[degrees]} & \colhead{[degrees]} & \colhead{} & \colhead{} & \colhead{[s]} & \colhead{[$\arcsec$]}
}
\startdata
SDSS\_J0912+1523	& 138.17821 & 15.38479 & 2016.1.01126.S & 2017-01-08 & 5866.56 & 1.57 \\
SDSS\_J2202-0033	 & 330.60121	 & -0.55955 & 2016.1.01126.S & 2017-03-07 & 5685.12 & 2.00 \\
SDSS\_J1448+1010 & 222.19133 & 10.16960 & 2017.1.01109.S & 2018-03-12 & 5999.616 & 0.74 \\
SDSS\_J0753+2403 & 118.43406 & 24.06005 & 2017.1.01109.S & 2018-03-20 & 5987.52 & 0.72 \\
SDSS\_J1203+1807 & 180.98548 & 18.13016 & 2017.1.01109.S & 2018-03-14 & 5927.04 & 0.70 \\
SDSS\_J1007+2330 & 151.80432 & 23.51530 & 2017.1.01109.S & 2018-03-15 & 5927.04 & 0.75 \\
SDSS\_J1053+2342 & 163.44737 & 23.70956 & 2017.1.01109.S & 2018-03-16 & 5987.52 & 0.77 \\
SDSS\_J0233+0052 & 38.49722 & 0.87734 & 2017.1.01109.S & 2018-04-04 & 5987.52 & 1.35 \\
SDSS\_J1302+1043 & 195.70387 & 10.71748 & 2017.1.01109.S & 2018-03-21 & 5987.52 & 0.67 \\
SDSS\_J1109-0040 & 167.38393 & -0.66774 & 2017.1.01109.S & 2018-04-10 & 5927.04 & 1.11 \\
SDSS\_J0046-0147 & 11.66247 & -1.78856 & 2017.1.01109.S & 2018-04-15 & 5987.52 & 1.02 \\
SDSS\_J0027+0129\tablenotemark{a} & 6.85600 & 1.49942 & 2017.1.01109.S & 2018-04-10 & 5927.04 & 1.11 \\
SDSS\_J2258+2313 & 344.52365 & 23.22115 & 2017.1.01109.S & 2018-05-01 & 5987.52 & 1.17 \\
\enddata
\tablenotetext{a}{This galaxy was also observed as part of 2016.1.01126.S, but those observations did not pass quality assessment (QA).}
\end{deluxetable*}

\subsection{ALMA CO(2--1) Observations}

We have targeted a subset of the \squiggle sample for extensive multi-wavelength follow-up studies, preliminary results of which have been published in several articles. In \citet{suess:17} we published ALMA CO(2--1)-based detections of vast $H_2$ reservoirs ($\log \Mgas/\msun\gtrsim10.0$, or $f_{H_2}\sim 20$\% and 4\%)\footnote{We note that in \citet{suess:17} we used delayed tau SFHs; the stellar masses based on flexible SFHs yield higher $M_{\star}$ values and therefore lower \fgas\ measurements of $\sim$14.5\% and $\sim$1\% for the same galaxies.} in two galaxies. These results suggest that the cold molecular gas is common in massive, recently quenched galaxies at $z\sim0.6$, but are far from conclusive. In the current paper, we present the demographics of $H_2$ (as probed by CO(2--1)) in a larger sample of 13 galaxies, adding 11 galaxies to those presented in \citet{suess:17}. The targeted subset is indicated by the green circles and histograms in Figures \ref{fig:distributions} and \ref{fig:hd_d4000_select}. We note specifically that the subset of galaxies selected for ALMA follow-up (green points) span the range of stellar populations of the full \squiggle\, sample in this parameter space. Properties of the observations including program numbers, observation dates, integration times, and spatial resolution of the data are included in Table \ref{tbl:alma} and physical properties of the sample are also enumerated in Table \ref{tbl:properties}. Because we chose this subsample for follow-up observations, our selection is biased towards brighter galaxies (in the i band) that can be observed by ALMA in the southern hemisphere. These targets are brighter due to a combination of slightly lower redshifts and higher masses than the full \squiggle selection. However, we note that the spectral diversity of the full \squiggle dataset, e.g., as probed by H$\delta_A$ and $D_n4000$ in Figure \ref{fig:hd_d4000_select}, is well-sampled by the ALMA targets.

Following the strategy of our pilot study \citep[ALMA Program \#2016.1.01126.S, PI: Bezanson]{suess:17}, we targeted CO(2--1) in 11 additional galaxies using ALMA band 4 in Cycle 5 (PI: Bezanson, Program\# 2017.1.01109.S). All targets were observed in two 80 minute observing blocks with the full 12m array. Total on-source integration times were ${\sim}1.7$ hours and the angular resolution of the resulting datacubes ranges from 0.7--2\arcsec (see Table \ref{tbl:alma}).

\begin{figure*}[t]
    \centering
    \includegraphics[width=0.48\textwidth]{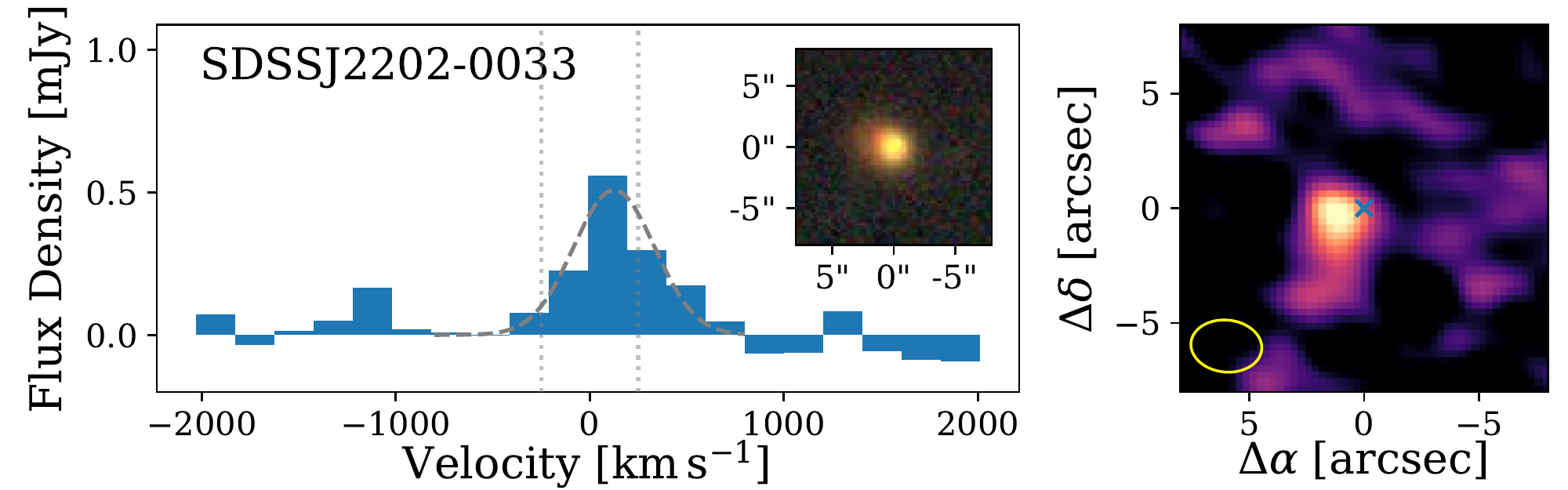}
    \includegraphics[width=0.48\textwidth]{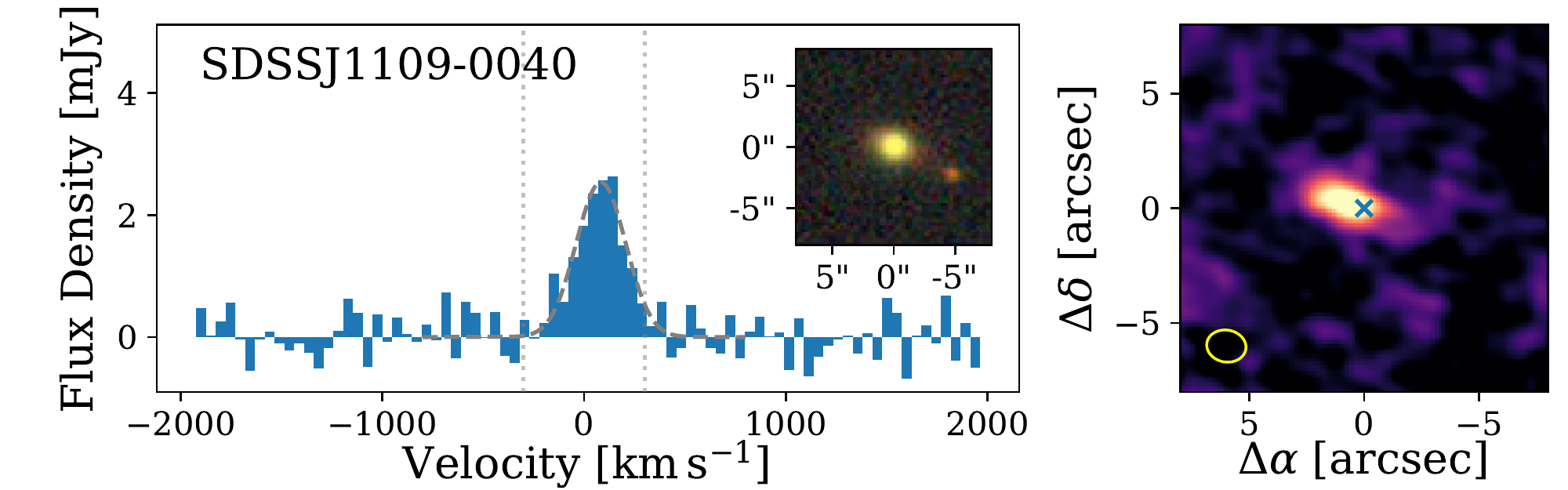}
    \includegraphics[width=0.48\textwidth]{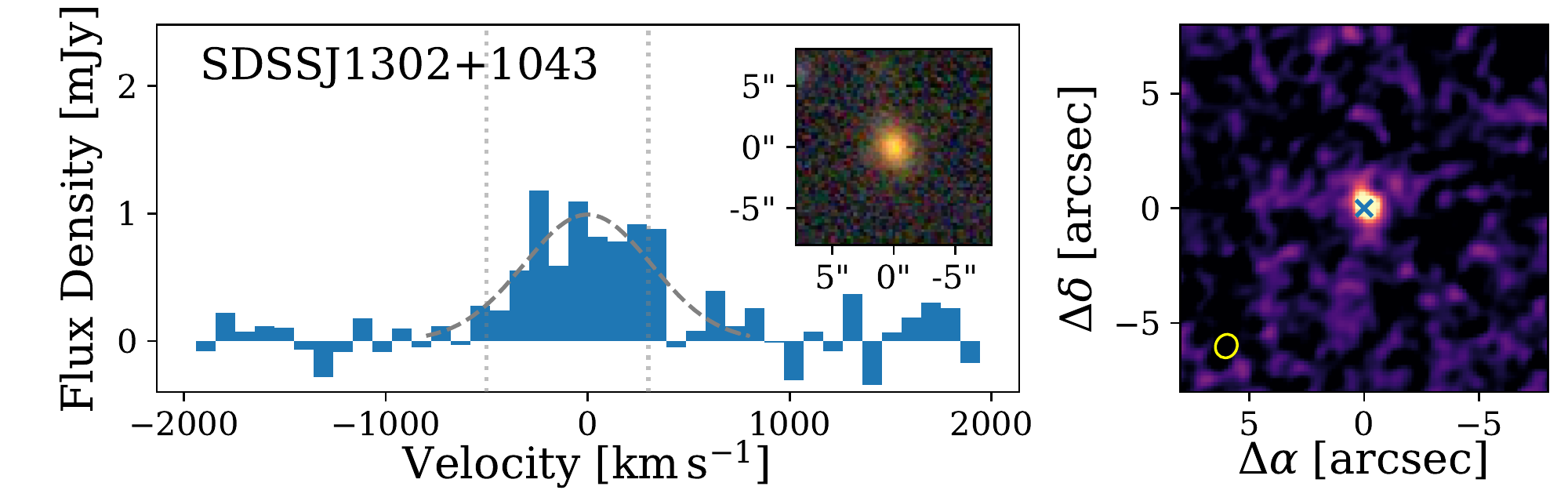}
    \includegraphics[width=0.48\textwidth]{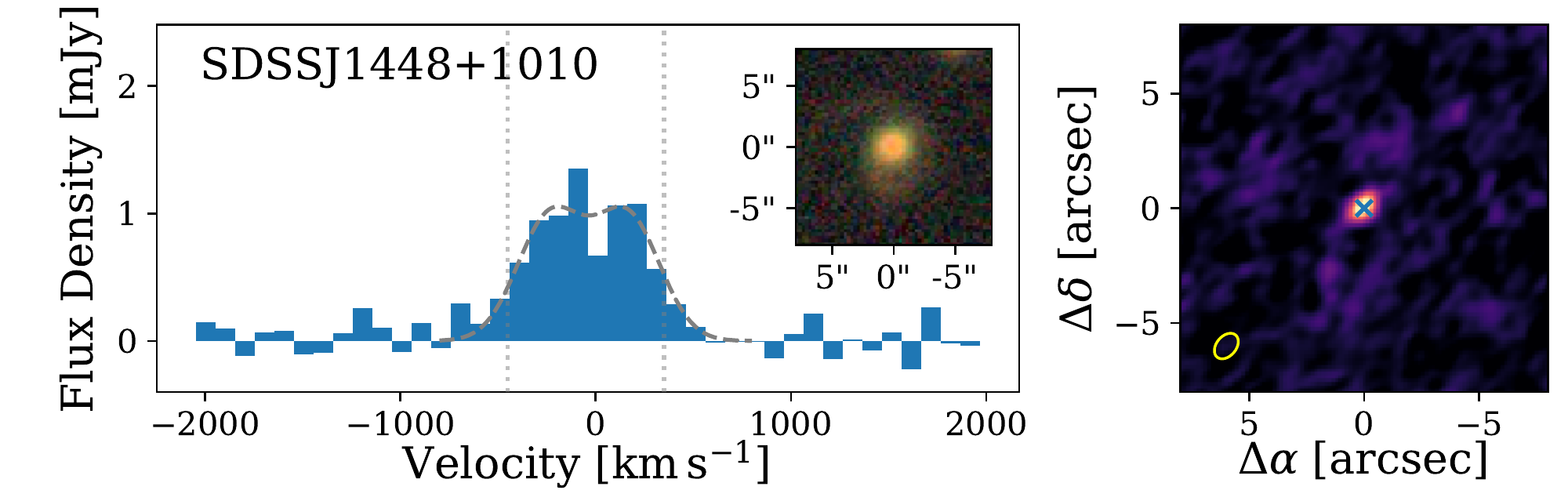}
    \includegraphics[width=0.48\textwidth]{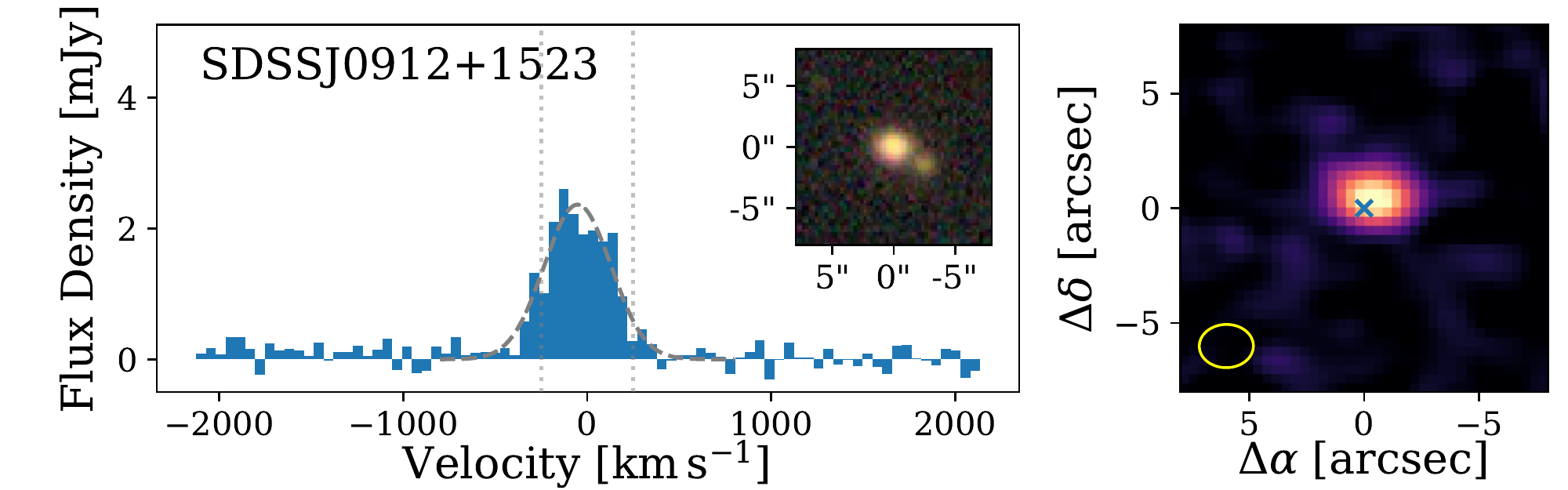}
    \includegraphics[width=0.48\textwidth]{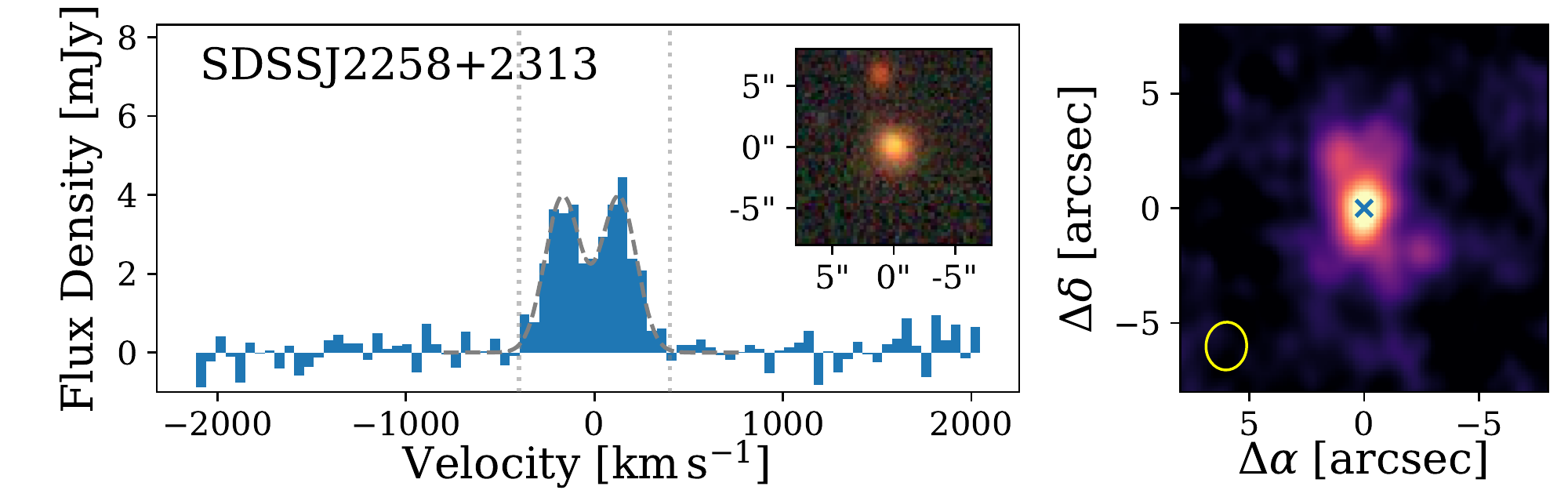}
    \caption{Spatially integrated CO(2--1) spectra (left) and line maps (right) for \squiggle galaxies that are detected in our ALMA band 4 follow-up study. { Gaussian spectral fits are included as dashed lines and line maps are collapsed between vertical dashed lines.} Galaxies are sorted from low to high $L_{CO(2-1)}$. { Inset panels show optical {\it grz} images from the DESI Legacy Survey DR9 \citep{dey:19}.} 
    \label{fig:COdetected}}
    
    \vspace{+20pt}
    \includegraphics[width=0.48\textwidth]{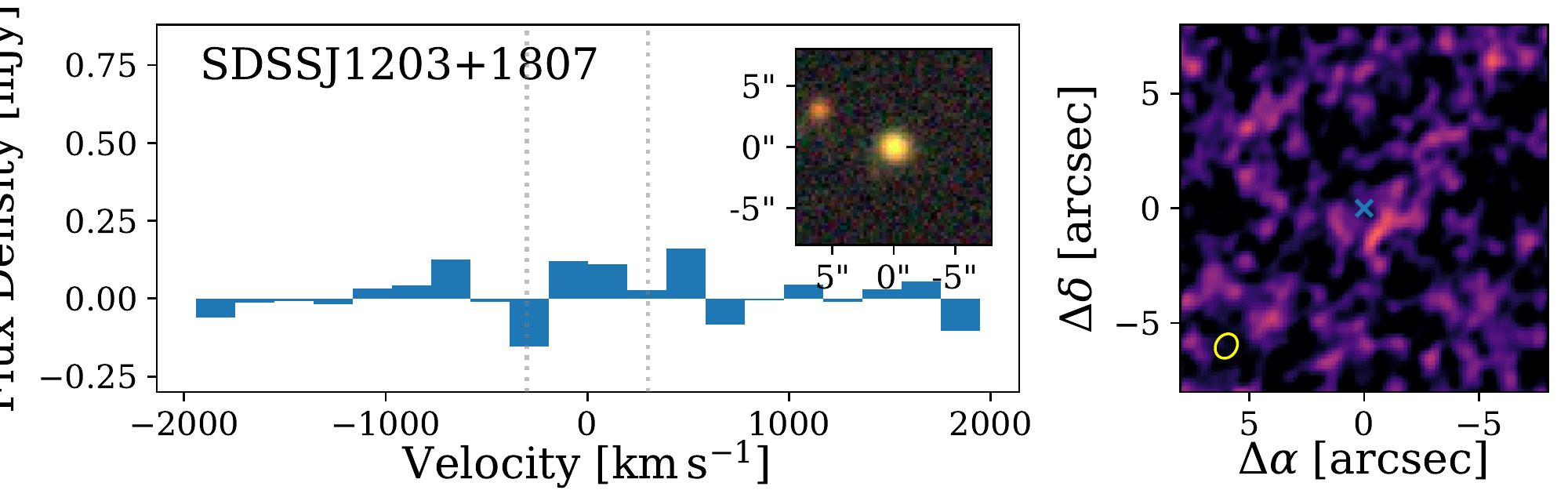}
    \includegraphics[width=0.48\textwidth]{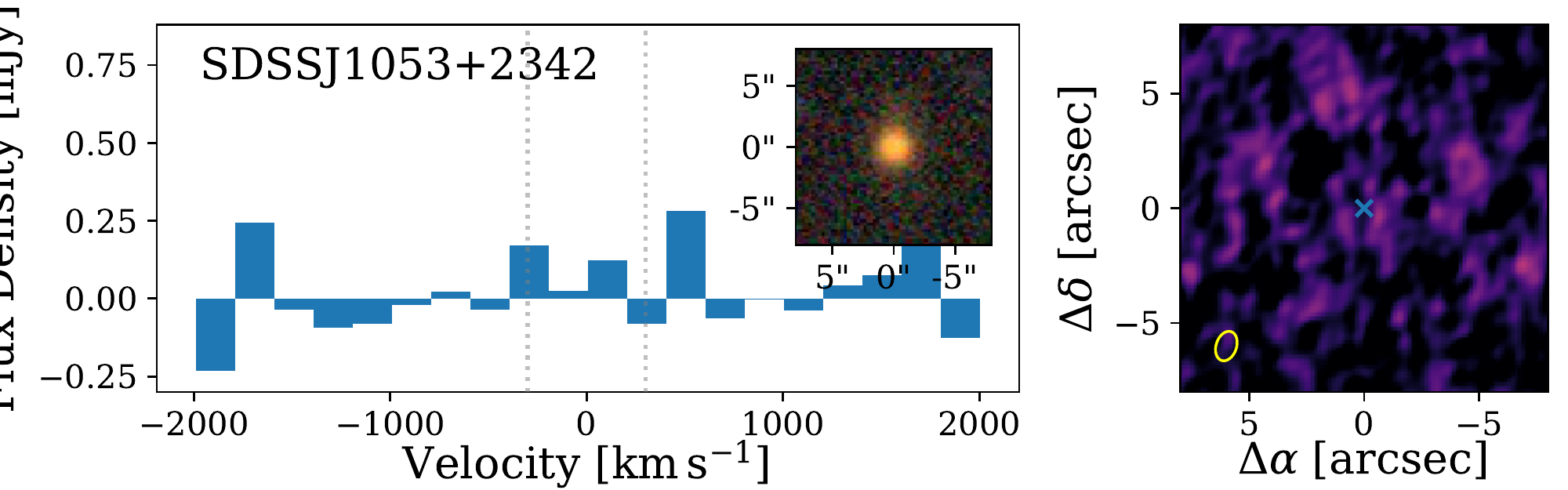}
    \includegraphics[width=0.48\textwidth]{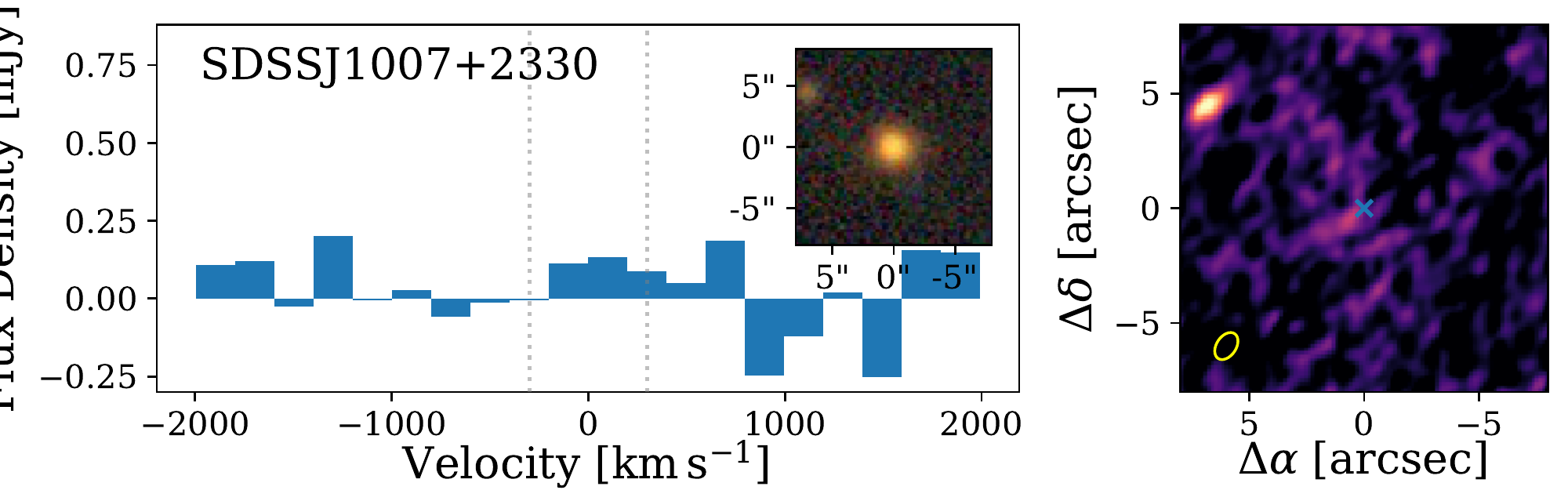}
    \includegraphics[width=0.48\textwidth]{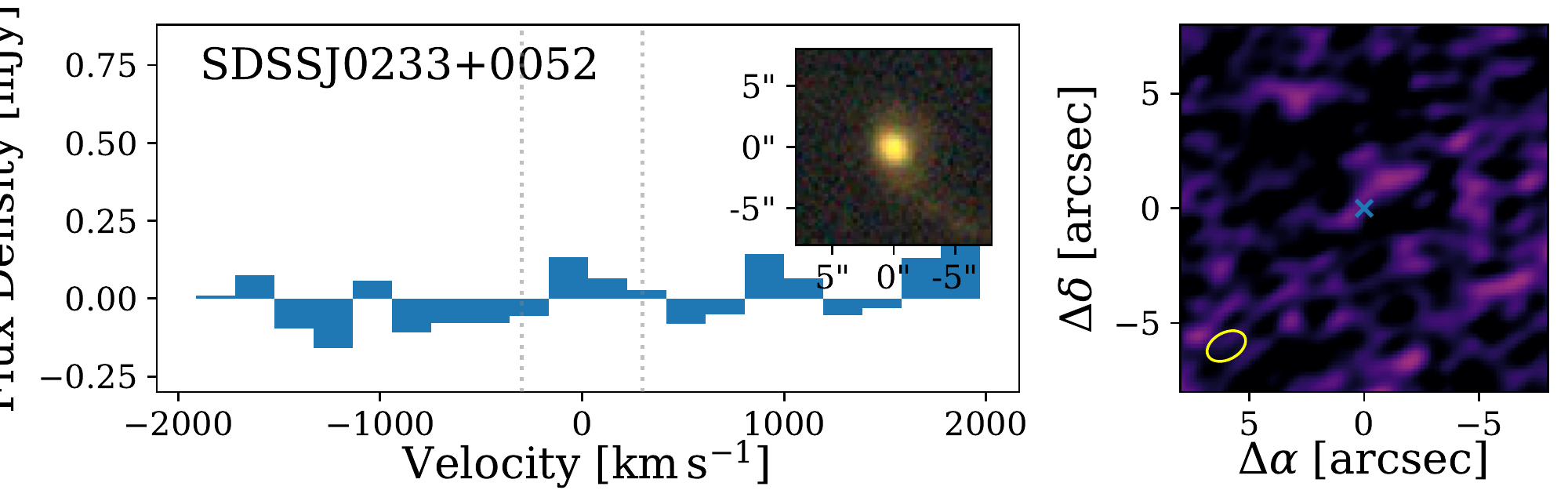}
    \includegraphics[width=0.48\textwidth]{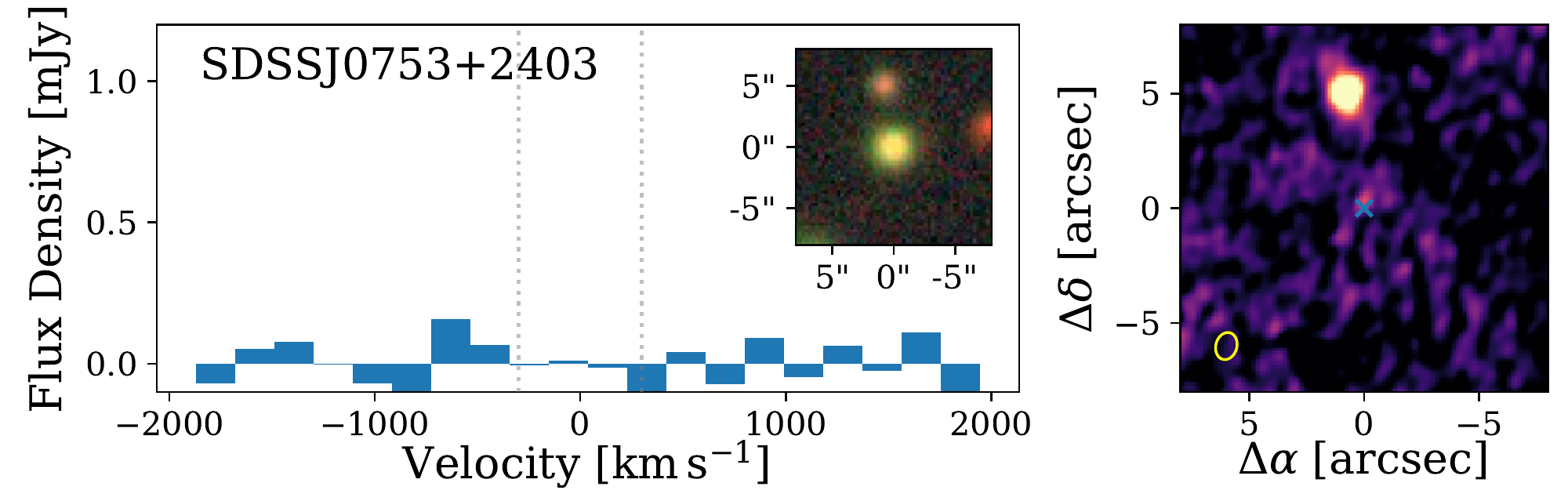}
    \includegraphics[width=0.48\textwidth]{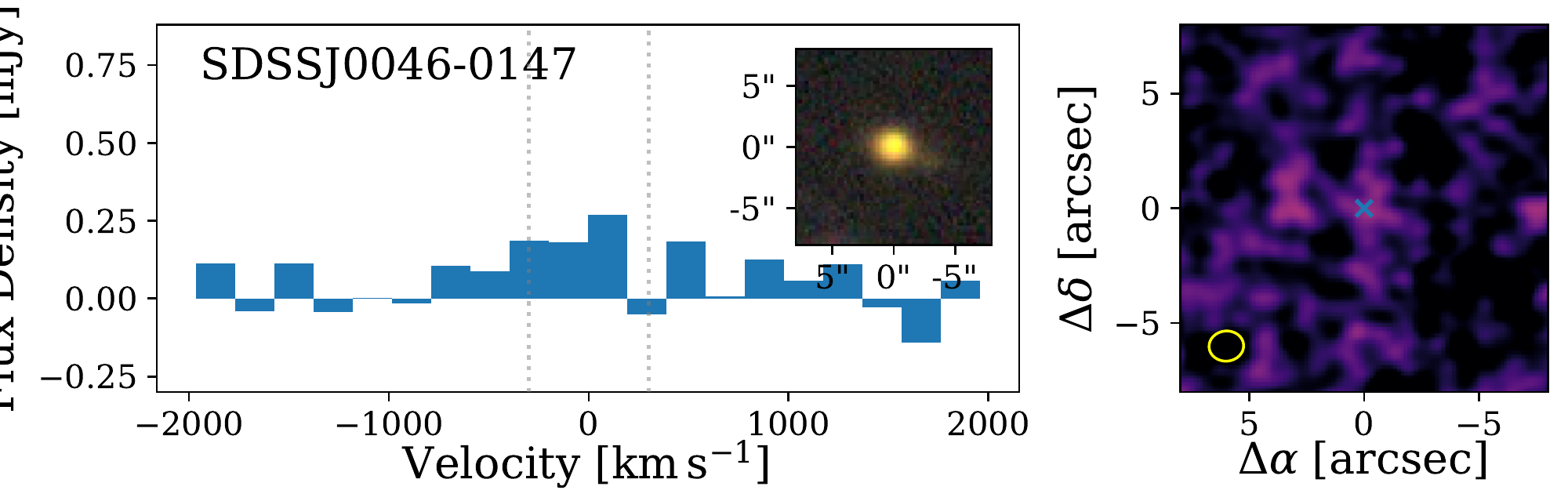}
    \includegraphics[width=0.48\textwidth]{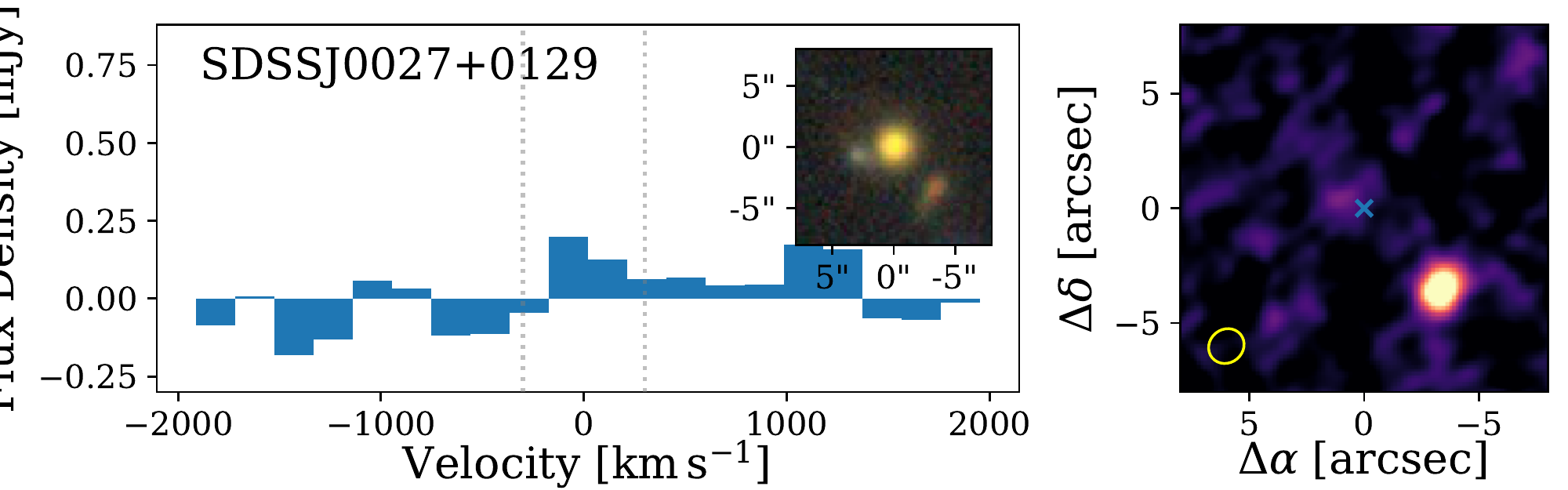}
    \includegraphics[width=0.48\textwidth]{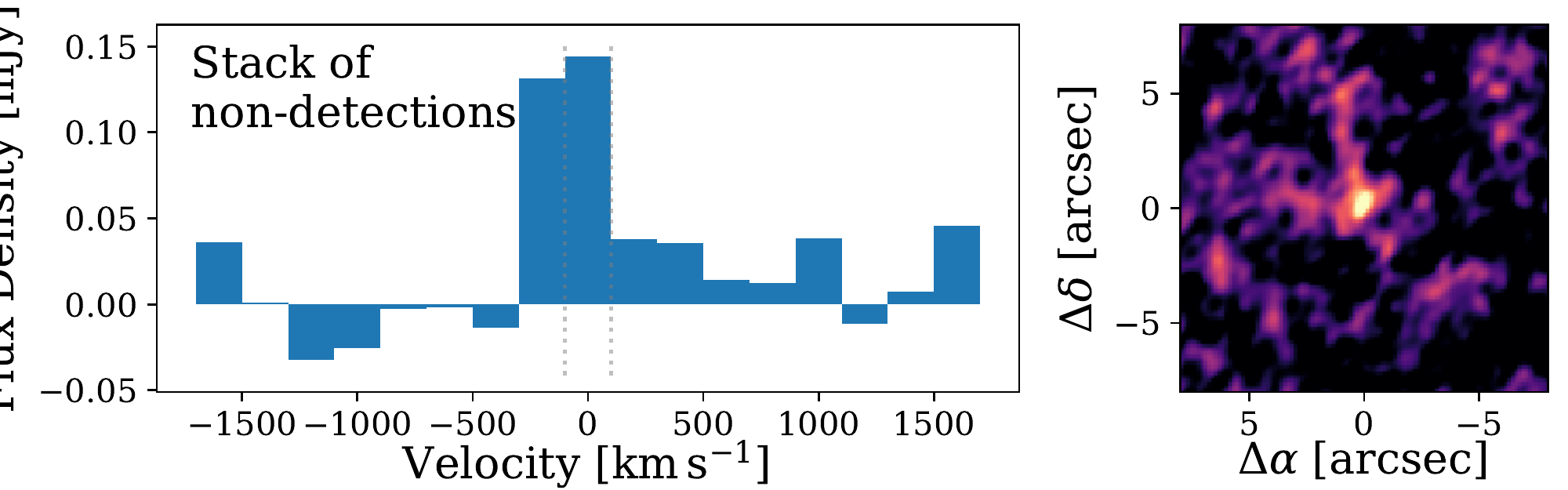}
    \caption{Spatially integrated CO(2--1) spectra (left) and line maps (right) { and optical images (left inset)} for undetected \squiggle galaxies, sorted by decreasing i magnitude. Although individual galaxies are not detected, the stacked  emission is detected (lower right). We note that we subtract a constant continuum offset from the spectrum of SDSS\_J0753+2403, which we attribute to non-thermal continuum emission.
    \label{fig:COundetected}}
\end{figure*}
\setlength{\tabcolsep}{2pt}

\begin{deluxetable*}{cccccccccccc}
\tabletypesize{\scriptsize}
\tablecaption{ALMA Target Sample Properties\label{tbl:properties}}
\tablehead{
\colhead{Galaxy ID} & \colhead{Plate-MJD-Fiber} & \colhead{$\mathrm{z_{spec}}$} & \colhead{$\mathrm{\log M_{\star}/M_{\odot}}$} & \colhead{SFR} & \colhead{Lick H$\delta_A$} & \colhead{$\mathrm{D_n4000}$} & \colhead{$\mathrm{SdvCO(2-1)}$} & \colhead{$\mathrm{L'_{CO(2-1)}}$} & \colhead{$\mathrm{R_{CO}}$} & \colhead{$\mathrm{\log M_{H_2}/M_{\odot}}$} \\ 
\colhead{} & \colhead{} & \colhead{} & \colhead{} & \colhead{[$M_{\odot}/yr$]} & \colhead{[$\mathrm{\AA}$]} & \colhead{} & \colhead{[$\mathrm{Jy\,km/s}$]} & \colhead{[$\mathrm{10^9\,K\,km/s\,pc^2}$]} & \colhead{[kpc]} & \colhead{}
}
\startdata
SDSS\_J1448+1010 & 5475-56011-379 & 0.6462 & 11.60$_{-0.07}^{+0.04}$& 1.06$_{-0.94}^{+0.99}$ &7.28$\pm$0.34 & 1.32$\pm$0.02 & 0.82$\pm$0.05 & 4.82$\pm$0.30 & 1.5$\pm$0.6 & 10.29$\pm$0.03 \\
SDSS\_J0753+2403 & 4466-55857-198 & 0.5652 & 11.32$_{-0.05}^{+0.03}$& 0.10$_{-0.09}^{+0.25}$ &8.99$\pm$0.21 & 1.32$\pm$0.01 & $<$0.09 & $<$0.42 &  \nodata & $<$9.22 \\
SDSS\_J1053+2342 & 6417-56308-55 & 0.6370 & 11.62$_{-0.08}^{+0.03}$& 0.29$_{-0.28}^{+0.67}$ &6.01$\pm$0.44 & 1.41$\pm$0.02 & $<$0.15 & $<$0.86 &  \nodata & $<$9.54 \\
SDSS\_J0027+0129 & 4302-55531-482 & 0.5851 & 11.53$_{-0.03}^{+0.03}$& 1.44$_{-1.15}^{+0.66}$ &8.23$\pm$0.34 & 1.29$\pm$0.02 & $<$0.12 & $<$0.59 &  \nodata & $<$9.38 \\
SDSS\_J2202-0033 & 1105-52937-311 & 0.6573 & 11.73$_{-0.02}^{+0.03}$& 1.99$_{-1.70}^{+1.91}$ &8.68$\pm$0.27 & 1.30$\pm$0.01 & 0.27$\pm$0.03 & 1.62$\pm$0.21 & $<$8.7 & 9.81$\pm$0.06 \\
SDSS\_J2258+2313 & 6308-56215-977 & 0.7058 & 11.82$_{-0.05}^{+0.03}$& 0.94$_{-0.71}^{+1.85}$ &6.21$\pm$0.53 & 1.30$\pm$0.02 & 1.86$\pm$0.08 & 13.10$\pm$0.56 & 12.4$\pm$0.9 & 10.72$\pm$0.02 \\
SDSS\_J0233+0052 & 705-52200-614 & 0.5918 & 11.61$_{-0.04}^{+0.02}$& 0.01$_{-0.01}^{+0.22}$ &8.39$\pm$0.41 & 1.42$\pm$0.02 & $<$0.11 & $<$0.56 &  \nodata & $<$9.35 \\
SDSS\_J0046-0147 & 4370-55534-762 & 0.6088 & 11.55$_{-0.03}^{+0.03}$& 0.14$_{-0.14}^{+1.11}$ &9.76$\pm$0.30 & 1.38$\pm$0.02 & $<$0.12 & $<$0.61 &  \nodata & $<$9.39 \\
SDSS\_J1109-0040 & 278-51900-193 & 0.5935 & 11.29$_{-0.03}^{+0.09}$& 2.33$_{-1.62}^{+1.12}$ &7.97$\pm$0.36 & 1.31$\pm$0.02 & 0.78$\pm$0.06 & 3.85$\pm$0.30 & 5.7$\pm$0.9 & 10.19$\pm$0.03 \\
SDSS\_J1203+1807 & 2595-54207-459 & 0.5946 & 11.44$_{-0.05}^{+0.02}$& 0.02$_{-0.02}^{+0.19}$ &7.07$\pm$0.42 & 1.46$\pm$0.02 & $<$0.09 & $<$0.45 &  \nodata & $<$9.26 \\
SDSS\_J1007+2330 & 6458-56274-501 & 0.6353 & 11.60$_{-0.03}^{+0.03}$& 0.89$_{-0.86}^{+0.99}$ &5.15$\pm$0.54 & 1.46$\pm$0.03 & $<$0.15 & $<$0.85 &  \nodata & $<$9.53 \\
SDSS\_J0912+1523 & 2438-54056-396 & 0.7473 & 11.37$_{-0.02}^{+0.03}$& 0.81$_{-0.76}^{+1.33}$ &9.03$\pm$0.33 & 1.24$\pm$0.01 & 1.07$\pm$0.05 & 8.47$\pm$0.36 & 6.2$\pm$0.8 & 10.53$\pm$0.02 \\
SDSS\_J1302+1043\tablenotemark{a} & 5421-55980-512 & 0.5921 & 11.61$_{-0.06}^{+0.04}$\tablenotemark{b} & 0.26$_{-0.26}^{+0.93}$ &6.18$\pm$0.93 & 1.38$\pm$0.04 & 0.78$\pm$0.09 & 3.85$\pm$0.42 & 3.3$\pm$0.7 & 10.19$\pm$0.05 \\
\enddata
\tablecomments{Upper limits for the undetected sources are 3$\sigma$, and assume an 800\,\kms line width.}
\tablenotetext{a}{This target was included in an early generation of the \squiggle sample based on the SDSS DR12 spectra, but the spectral shape differs slightly in the DR14 spectrum, causing the rest-frame colors to fall out of the \squiggle selection cuts. { We include this galaxy in all further analysis.}}
\tablenotetext{b}{The \texttt{Prospector} modeling for this galaxy is unable to reproduce the WISE $4.5\mu$m flux, driving the fit against the edges of the priors for dust. We expect that this is due to AGN, AGB, or neighboring light that is not accounted for in the model parameter space. Therefore, we exclude WISE band 2 in for this object and the fit converges well.}
\end{deluxetable*}
\begin{figure*}[!t]
    \centering
    \includegraphics[width=\textwidth]{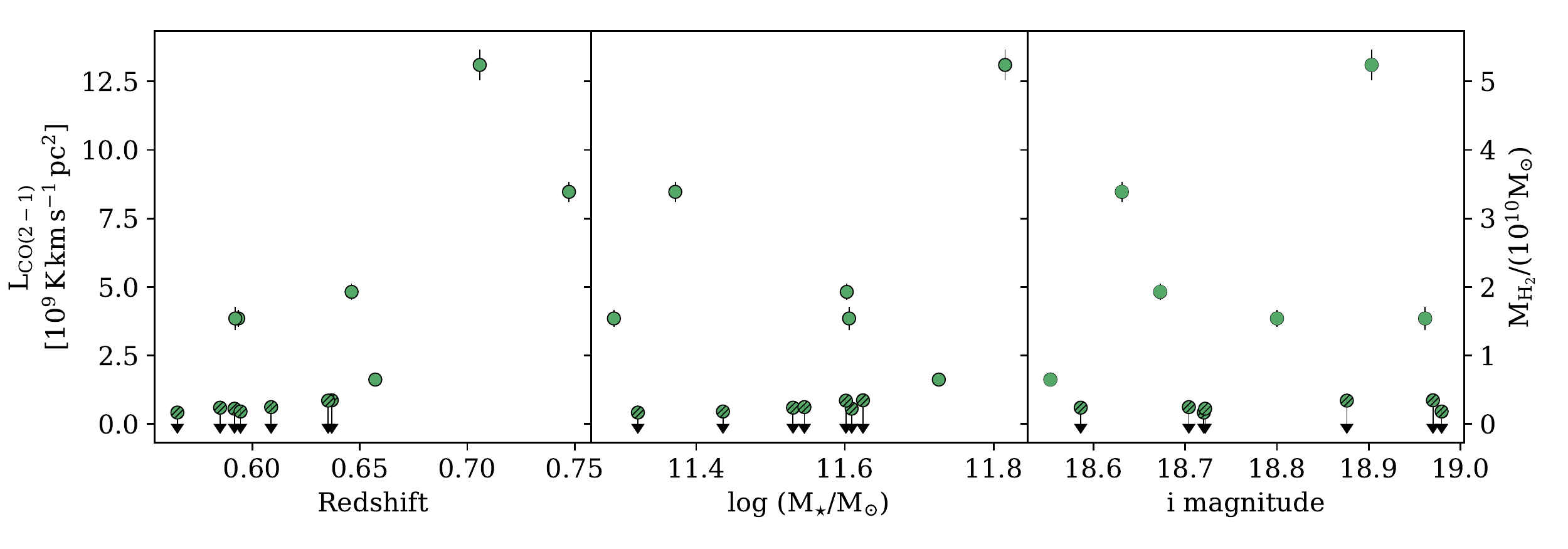}
    \caption{CO(2-1) Luminosity (left axis) and $M_{H_2}$ (right axis, assuming $\alpha_{CO}=4.0$) versus redshift (left panel), stellar mass (center panel), and i magnitude (right panel). Galaxies are primarily detected in the higher redshift subsample ($z\gtrsim 0.65$), but are otherwise detected across the full range of stellar mass and magnitude.}
    \label{fig:LCO}
\end{figure*}

Continuum images at $\sim$2\,mm were created using the full bandwidth of the ALMA data, excluding $\pm$500\,\kms\ around the expected frequency of the redshifted CO(2--1) line. The data reach continuum sensitivities $\sim$6-10\,$\mu$Jy. Continuum emission was detected in one target, SDSS\_J0753+2403; the flux ratio between the upper and lower sidebands of the ALMA data and a detection at 1.4GHz \citep{greene:20} make clear that this is non-thermal synchrotron emission from AGN activity. Following \citet{suess:17}, the non-detections of the other sources imply upper limits on the presence of any highly obscured star formation, SFR $\lesssim$ 50\,\msun/yr, assuming a standard modified blackbody function and a dust temperature $T_{\rm{dust}}=30$\,K. These limits are highly uncertain due to the unknown dust temperature because observed-frame 2\,mm is far from the peak of the dust SED, but are sufficient to rule out very high obscured SFRs.

All CO(2--1) spectra are extracted by fitting circular Gaussian models to the visibilities using \texttt{uvmultifit} \citep{martividal:14}. We generate spectra at effective velocity resolutions of $\approx$50, 100, and 200\,\kms. In the sources where CO is clearly detected, we often find evidence of velocity gradients and/or marginally-resolved source sizes in individual channels. We allow the position and size of the spatial Gaussian components to be free parameters across the line profiles, and fix them to the phase center and the median size for line-free channels. In undetected sources, we simply fix the centroid to the phase center and the size to 2'' in all cases, a value typical of the detected sources.

Integrated line fluxes are estimated by fitting extracted spectra, using one (spectral) Gaussian 

\noindent (SDSS\_J2202-0033, SDSS\_J1109-0400, SDSS\_J13202+1043, and SDSS\_J0912+1523) or two Gaussians (SDSS\_J1488-1010 and SDSS\_J2258-2312) to the CO(2--1) line profiles.  For undetected galaxies, the upper limits on CO(2--1) line flux are conservatively estimated using a single 800 \kms\ wide channel; upper limits on the line fluxes for these sources scale as $\sqrt{\Delta V / 800{\rm \kms}}$ for alternative choices of the velocity width $\Delta V$. We stack the integrated spectra of the seven galaxies that are individually undetected using the 200\,\kms spectra, finding a $\approx 3.7\sigma$ detection of CO(2--1) in the stacked spectrum. We verify this detection with a simple image-plane stack of the CO data cubes, but stress that this is merely for visualization purposes due to the varying spatial resolution of the input data cubes. We estimate the spatial extent of CO(2--1) emission in each detected galaxy by fitting 2D Gaussians in the image plane. Circularized  half-width-half-max (HWHM) values are quoted in Table \ref{tbl:properties} in physical units (kpc).

Of the 13 targeted galaxies, CO(2--1) emission was detected in six. The CO(2--1) spatially integrated spectra and line maps for the detected \squiggle galaxies are included in Figure \ref{fig:COdetected} and undetected galaxies are included in Figure \ref{fig:COundetected}. { Spectral Gaussian fits are included as dashed lines and vertical lines indicate the regions used to generate the linemaps.}{ Optical images (grz) are included from the DESI Legacy Survey \citep{dey:19} as insets to the CO(2--1) spectra.} We note that of the seven galaxies that are undetected in CO(2--1), three datacubes include significantly detected lines in close physical and kinematic proximity to the \squiggle galaxies. We interpret these as representing CO(2--1) emission from the cold gas reservoirs of neighboring galaxies. We note that due to the high stellar masses of the galaxies in this sample, we expect them to reside in dense environments, but defer the analysis of these nearby sources to future work. We detect non-thermal continuum emission in SDSS\_J0753+2403, which we subtract from the spectrum in Figure \ref{fig:COundetected}.

Of the 6 detected galaxies, CO(2--1) is largely spatially coincident with the optical centroid of the galaxies, as determined from the SDSS imaging, with the exception of SDSS\_J1109-0040 and SDSS\_J2202-0033 \citep[also in ][]{suess:17}. In SDSS\_J2202-0033 the $\sim$1'' offset from the optical centroid is not significant given the resolution and signal-to-noise of the data{; we note that the optical image of this galaxy appears to be slightly asymmetric}. The CO(2--1) emission in SDSS\_J1109-0040 is spatially offset by $\sim1-3''$; it is possible that this galaxy is in the process of a close, late-stage merger for which only one galaxy is detected in CO. { A possible companion galaxy is visible in the optical image, on the opposite side than the direction of elongation}. Alternatively we may be detecting an especially strong CO outflow. In either case the emission is not especially broad ($\sigma=123$ \kms) in comparison to the other detected sources. In two additional sources, SDSS\_J1448+1010 and SDSS\_J2258+2313, the CO emission appears to extend beyond the optical extent of the galaxies, reaching distances up to $\sim$4-5'' from the center. These two objects may similarly be exhibiting late-stage mergers or molecular outflows in which a significant fraction of the molecular gas has been removed, or the gas may be associated with low surface-brightness stellar light not apparent in the SDSS imaging. These offset and extended sources are targets of follow-up ALMA observations that will be presented in future work. For this work we assume that all CO(2--1) emission detected is physically associated with the \squiggle\ post-starburst galaxies.

Figure \ref{fig:LCO} shows the CO(2--1) luminosity versus redshift (left), stellar mass (center), and i magnitude. To calculate $H_2$ masses from CO(2--1) we assume $r_{21}=1.0$, making the conservative assumption of thermalized emission, \citep[e.g.,][]{combes:07,dannerbauer:09,young:11} and a Milky Way-like CO-H$_2$ conversion factor of $\alpha_{CO}=4.0$ \citep[][and references within]{bolatto:13}. We note that these  assumptions are unlikely to lead to significant uncertainties in the implied $H_2$ properties. $r_{21}$ could only be lower than assumed, which would increase the $M_{H_2}$. Higher values of $\alpha_{CO}$ would require either cooler and lower velocity dispersion gas (increasing the CO line opacity) or lower metallicities (decreasing the CO abundance) than in the Milky Way, both of which are unlikely. Lower $\alpha_{CO}$ could arise from hotter gas or higher velocity dispersions than typical giant molecular clouds in the Milky Way, which may be realistic in this sample. However, a factor of $\sim2-3$ reduction in $\alpha_{CO}$ would be counteracted by any decrease in $r_{21}$ if the gas is not thermalized \cite[e.g.,][]{narayanan:12}. Implied $H_2$ masses are indicated on the right axis in this figure. In the left panel there is a weak trend indicating that the higher redshift galaxies in the \squiggle sample are more likely to be detected in CO(2--1), which is possibly suggestive of an evolutionary sequence, however we emphasize that the selection criteria are fairly narrow. CO(2--1) luminosity and \Mgas\, are independent of stellar mass or magnitude.

\begin{figure}
    \centering
    \includegraphics[width=0.49\textwidth]{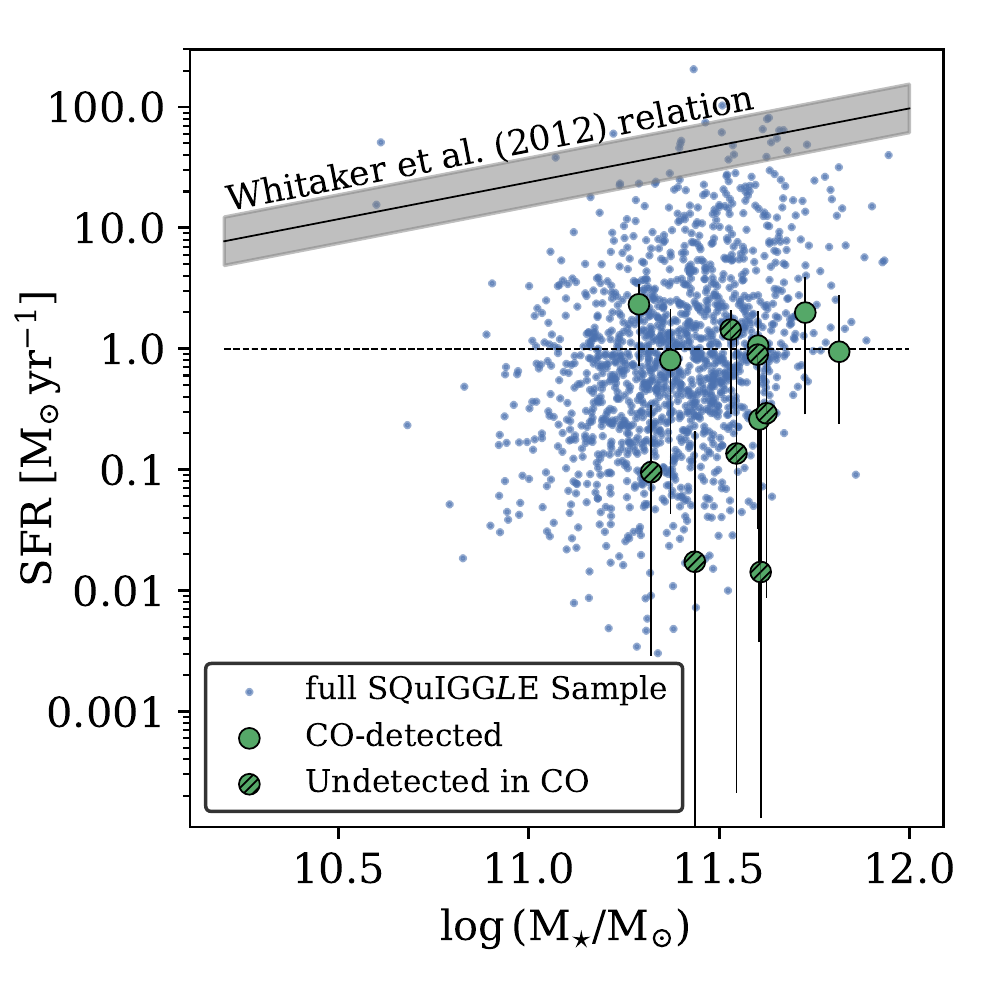}
    \caption{SFR versus \mstar\ measured from joint photometric and spectroscopic modelling using \texttt{Prospector}, for the full \squiggle sample (blue) and ALMA targets,  with the 1 M$_{\odot}$\,yr$^{-1}$ threshold below which SFRs are robustly recovered  (K. Suess, et al. in prep). Green circles indicate those that are detected in CO(2--1), hatched filling identifies undetected galaxies. We include the scaling relation for star-forming galaxies at the average redshift of the sample from \citet{whitaker:12b}. The \squiggle galaxies span a wide range of SFRs, but CO-detected targets fall below the main sequence by at least an order of magnitude.}
    \label{fig:SFR_Mstar}
\end{figure}

\section{$H_2$ reservoirs of \squiggle galaxies} \label{sec:gas}

\subsection{$H_2$ and Scaling Relations}
In this section we compare the enigmatic $H_2$ reservoirs of \squiggle galaxies to the scaling relations defined by ``normal'' star-forming galaxies at high and low-redshift and to other samples of sub-main sequence galaxies. In Figure \ref{fig:SFR_Mstar}, we show the SFRs of the \squiggle galaxies in blue (full sample) and green symbols for the ALMA targets. We include the ``star-forming main sequence'' scaling relation for star-forming galaxies at the average redshift of \squiggle\ \citep{whitaker:12b}. The \squiggle\ sample lies significantly below this relation, although many have comparable gas reservoirs to coeval star forming galaxies. The SFRs for the CO-detected \squiggle galaxies (green circles) are on average systematically higher by a factor of $\sim4$ than the undetected counterparts (hatched circles). However, we emphasize that measuring very low SFRs (SFR$\lesssim1 M_{\odot}$yr$^{-1}$, dashed line) is extremely challenging.

\begin{figure*}[!t]
    \centering
    \includegraphics[width=0.45\textwidth]{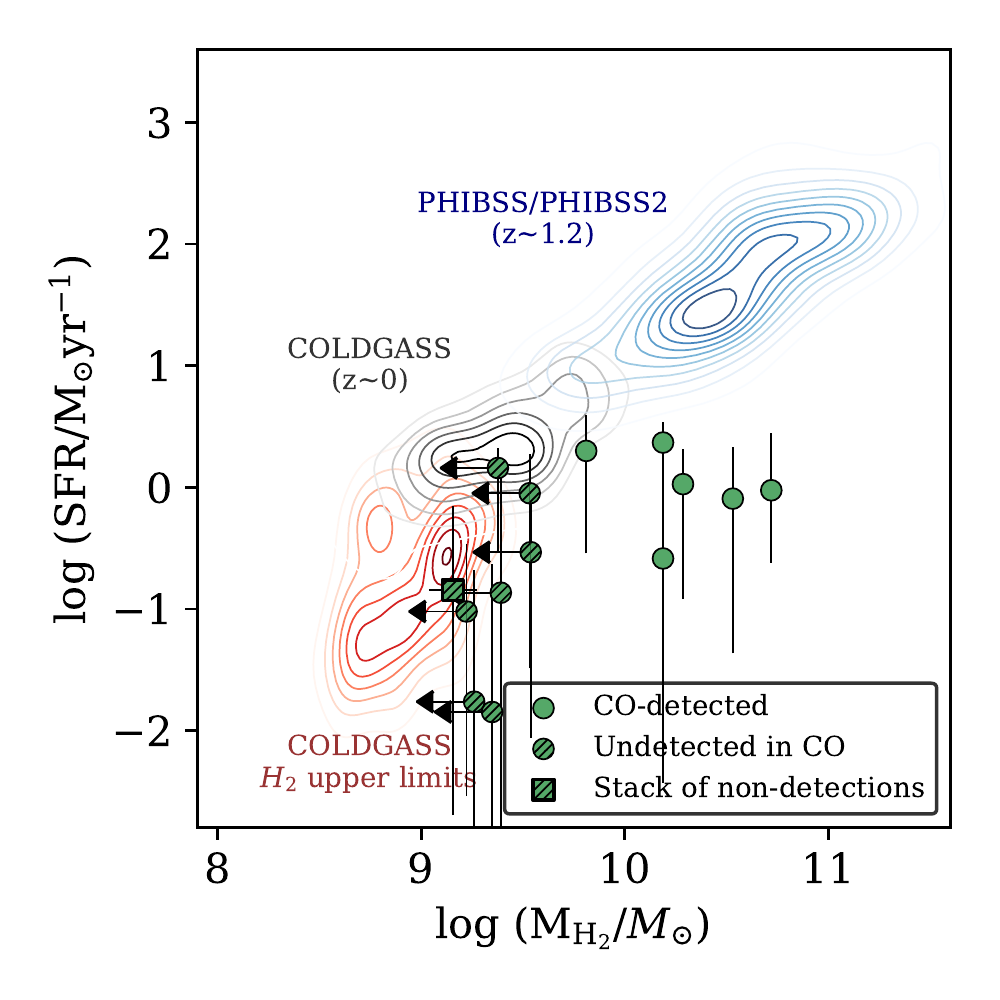}
    \includegraphics[width=0.45\textwidth]{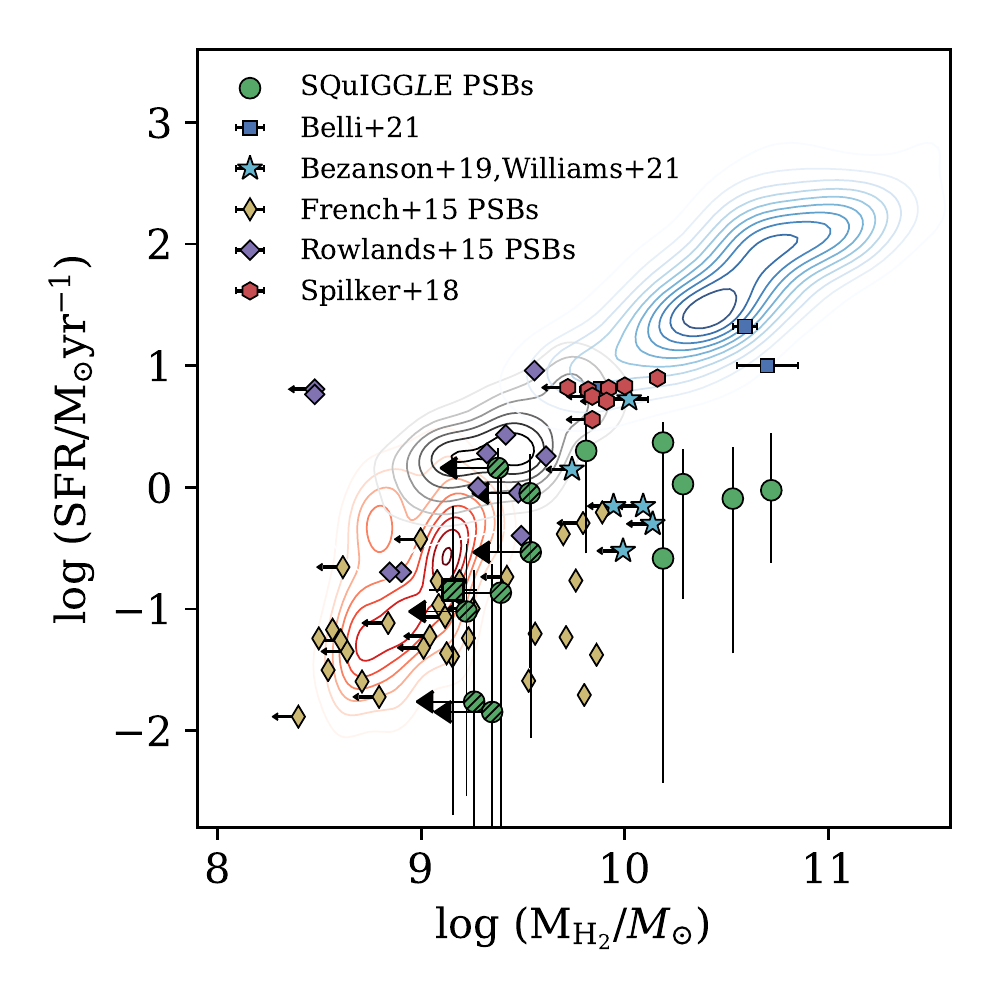}
    \caption{SFR versus $H_2$ gas mass for \squiggle galaxies and comparison samples of massive galaxies from COLDGASS at $z\sim0$ \citep{saintonge:11,saintonge:11b} and star-forming galaxies at $\langle z\rangle\sim1.2$ from PHIBSS/PHIBSS2 \citep{tacconi:10,tacconi:13}. Green symbols in both panels indicate the \squiggle post-starburst galaxies. Additional samples of $z\sim0$ post-starburst galaxies \citep{french:15,rowlands:15} and galaxies below the star-forming SFR-stellar mass ``main sequence'' with CO-based $H_2$ mass estimates are included in the right panel \citep{spilker:18,bezanson:19,williams:21, belli:21}. Although most galaxies lie near the \Mgas - SFR relation,  a number of \squiggle and \citet{french:15} post-starburst galaxies harbor large $H_2$ reservoirs for their low star formation rates. This is most dramatic for the CO(2--1)-detected \squiggle galaxies, which are offset by over an order of magnitude in \Mgas.}
    \label{fig:SFR_Mgas}
\end{figure*}
\begin{figure}[!h]
    \centering
    \includegraphics[width=0.45\textwidth]{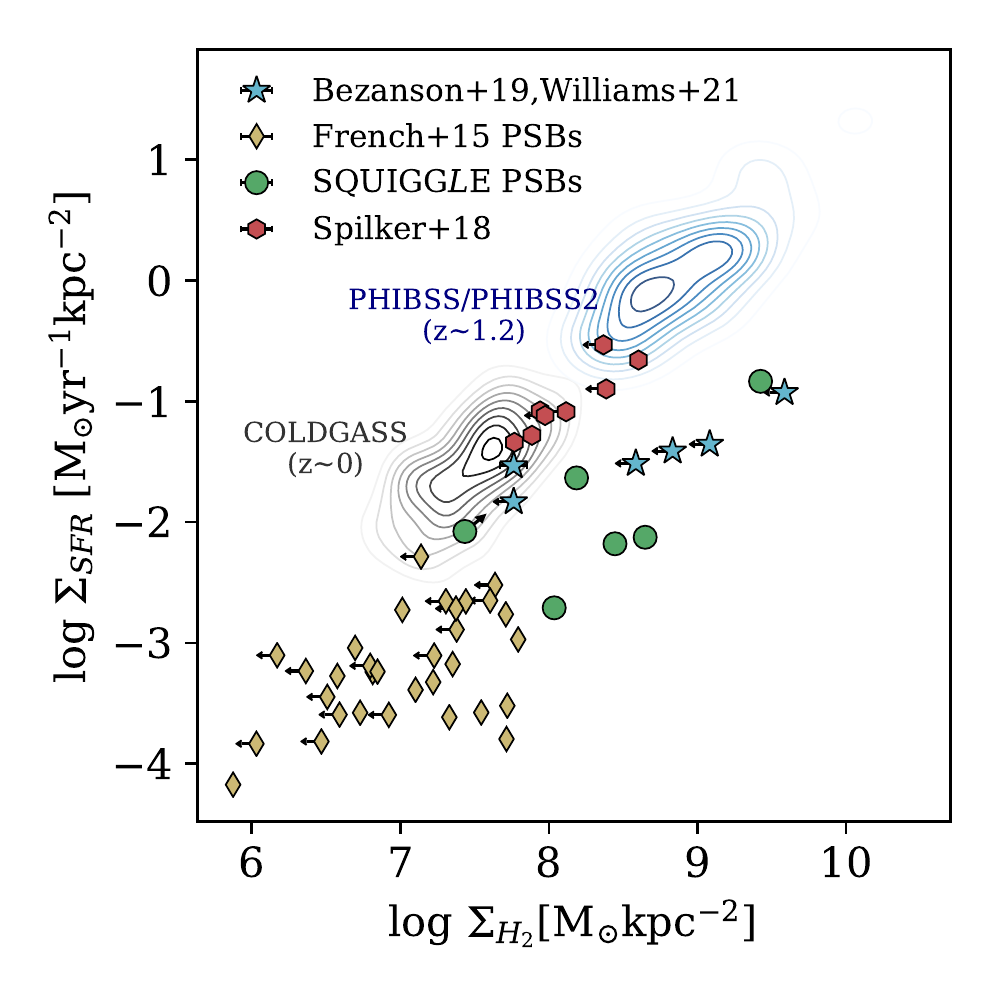}
    \caption{SFR surface density versus $H_2$ surface density, or the Schmidt-Kennicutt relation for the detected \squiggle galaxies. Sizes are estimated from the CO(2--1) emission, which is likely an overestimate of the stellar sizes. Even at the most extended limit, this sample lies at the highest density edge of the local \citet{french:15} post-starburst galaxies and more than an order of magnitude offset from the K-S relation for ``normal'' star-forming galaxies.}
    \label{fig:KS}
\end{figure}

Figure \ref{fig:SFR_Mgas} shows the SFR versus $\mathrm{H_2}$ gas mass for the \squiggle post-starburst galaxies (green symbols). The empirical scaling relation for CO-based $\mathrm{H_2}$ measurements is traced by colored contours. At lower SFR, massive galaxies at $z\sim0$ from COLDGASS \citep{saintonge:11,saintonge:11b,saintonge:12} are shown by black (for galaxies with detected CO(1--0) lines) and red (upper limits on CO(1--0)) contours. Star-forming galaxies at $\langle z \rangle=1.2$ from PHIBSS/PHIBSS2 surveys \citep{tacconi:10,genzel:15,freundlich:19} are shown by blue contours. In the left panel, only individual \squiggle galaxies and the \squiggle stack are included. Detected galaxies are offset by $\gtrsim 1$dex in $M_{H_2}$, while the stack of non-detections (hatched square) is consistent with the upper limits of galaxies with low SFRs in the local Universe. In the right panel we include other samples of galaxies with SFRs that place them below the main sequence of star-forming galaxies \citep[e.g.][]{noeske:07}. This panel includes post-starburst galaxies at $z\sim0$ from the SDSS \citep{french:15,rowlands:15}. Much like the galaxies in \squiggle, post-starburst galaxies from \citet{french:15} (yellow diamonds) exhibit a range of $H_2$ reservoirs, which places some galaxies $\sim0.5$dex more gas rich than the upper limits of local low-SFR galaxies. Post-starburst galaxies from \citet{rowlands:15} (purple diamonds) have higher quoted SFRs and lower \Mgas, lying well within the distribution of galaxies in the COLDGASS sample. At higher redshift we also include galaxies at a roughly coeval $\langle z \rangle=0.7$ from \citet{spilker:18} as maroon hexagons, which are selected from the LEGA-C survey \citep{wel:16} to be sub-main sequence, but have optical spectra that suggest more gradual star formation histories than post-starburst galaxies. Finally we include CO-based measurements of \Mgas\ for nine quiescent galaxies at $z\sim1.5$ (four detections and five upper limits), including two with post-starburst spectral signatures (strong Balmer absorption lines) \citep{bezanson:19, belli:21, williams:21}. The $H_2$ reservoirs of the galaxies detected by \citet{belli:21} (blue squares) are comparable to those of the \squiggle galaxies, however their higher SFRs (likely due to the fact that they are observed at higher redshift) place them closer to the coeval star-forming galaxies in the PHIBSS/PHIBSS2 survey \citep{tacconi:13,tacconi:18}.

It is useful to examine the SFR and $H_2$ masses in the more traditional projection of the Kennicutt-Schmidt relation \citep[e.g.][]{kennicutt:98}, as the density of cold gas is a more direct tracer of the fueling of star formation. As the galaxies are also unresolved in the existing ground-based imaging, we adopt circularized sizes derived from the CO(2--1) emission. Figure \ref{fig:KS} shows the $\Sigma_{SFR}$ versus $\Sigma_{H_2}$ for \squiggle galaxies that are detected in CO(2--1) and other samples with reliable size measurements, using the same symbols as Figure \ref{fig:SFR_Mgas}. We note that all other samples are included using sizes estimated from stellar effective radii. Given the uncertainty in spatial extent that impacts both surface densities, detected \squiggle galaxies can only come closer to the tight Kennicutt-Schmidt relation if their CO(2-1) emission is more extended than the spatial distribution of any residual star formation. 

\begin{figure}
    \centering
    \includegraphics[width=0.45\textwidth]{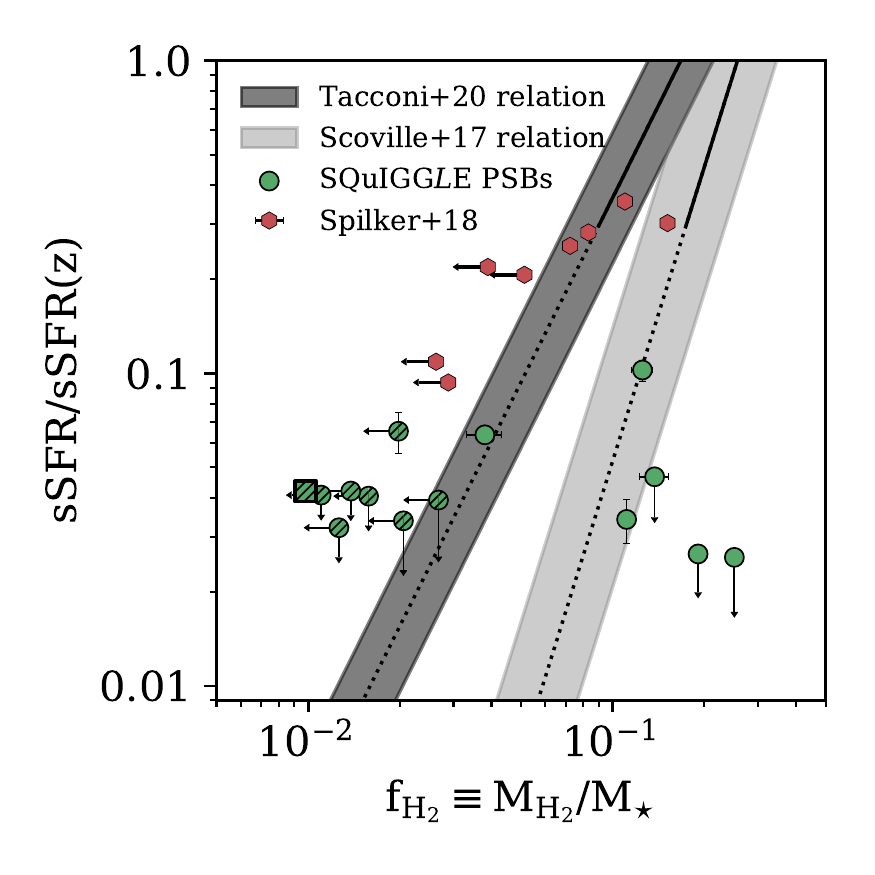}
    \caption{Specific star formation rate, normalized by the star-forming main sequence at the average redshift of the sample, versus $H_2$ gas fraction for quiescent galaxies from this sample and a coeval sample of less extreme quiescent galaxies from \citet{spilker:18}. The combined sample reveals a clear increase in the scatter about integrated scaling relations in the quiescent regime, even when accounting for lower efficiency of star formation in depleted galaxies.}
    \label{fig:ssfr_fgas}
\end{figure}

\begin{figure}
    \centering
    \includegraphics[width=0.45\textwidth]{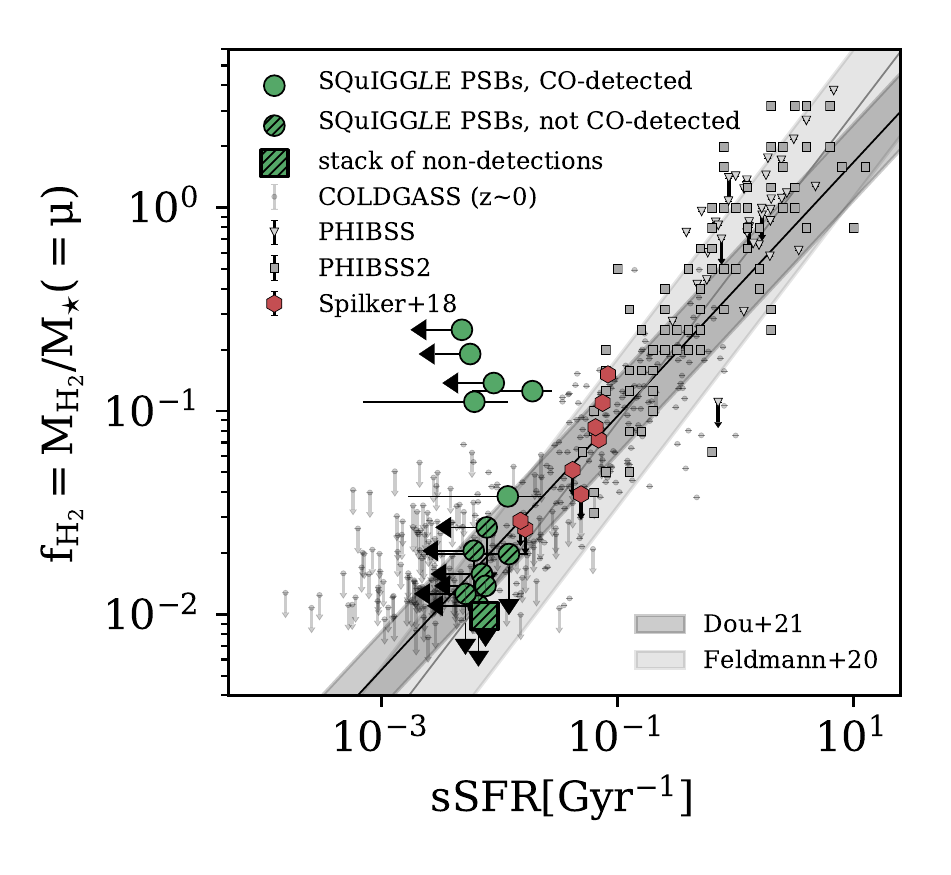}
    \caption{$H_2$ fraction versus sSFR in absolute quantities, representing a suggested fundamental relation within which galaxies evolve \citep[e.g.][]{feldmann:20,dou:21}. We note that the gas rich \squiggle galaxies lie significantly outside the scatter in this relation, but the deviation may be temporary and the stack of older, non-detections lies on the extended relation.}
    \label{fig:mu_ssfr}
\end{figure}

A number of groups have investigated additional integrated scaling relations between gas reservoirs and the fueling of star formation and star formation efficiency in galaxies \citep[e.g.][]{lilly:13,tacconi:13,scoville:17,tacconi:18,tacconi:20}. In part motivated by the time evolution of the characteristic SFR($M_{\star}$) of galaxies, some papers advocate for redshift-evolving multivariate regressions that minimize the scatter, quantifying e.g.\ $M_{H_2}(SFR, M_{\star}, z)$. The majority of the observational constraints on these scaling relations rely on galaxies that are near, on, or above the star-forming main sequence at any epoch. As a result, this sample provides a useful tool to probe the scatter about these relations at low SFRs. As motivated in \S2, we adopt slightly different definitions for stellar mass and SFR in Figures \ref{fig:ssfr_fgas} and \ref{fig:mu_ssfr}. These definitions are more consistent with those used to derive the scaling relations. We use $\mathrm{M_{\star,FAST}}$, from delayed exponential SFHs, for the former and treat SFR$=1 M_{\odot}$yr$^{-1}$ as a floor for derived values, below which SFRs are plotted as upper limits.

Figure \ref{fig:ssfr_fgas} shows the specific SFR (sSFR), normalized relative to the star forming main sequence, versus $H_2$ gas fractions for the sample of post-starburst galaxies presented in this paper. We also include a coeval sample of massive and sub-main sequence galaxies from \citet{spilker:18}. These individual measurements can be contrasted with extensions of two redshift-dependent scaling relations from \citet{scoville:17} (dark gray band) and \citet{tacconi:20} (light gray band), plotted assuming the average redshift and stellar mass of the \squiggle\ ALMA sample. For each scaling relation the quoted scatter is indicated by the gray band and the scaling relation is indicated by a solid gray or black line where the relation is calibrated and dotted lines to indicate where each is extrapolated. Although there is some agreement between the galaxies with higher sSFR and the scaling relations, at lower specific star formation rates, where the two relations diverge, it is clear that there is { a significant offset} between the $H_2$-detected \squiggle\ galaxies (green circles) and the undetected counterparts (hatched green symbols) - individually or in the stack (square). Detected galaxies would need to have \emph{vastly} underestimated star formation rates (by at least an order of magnitude) to be consistent with either scaling relation. Such high SFRs are disfavored by e.g., the non-detection of 2mm dust continuum emission in any source.

\begin{figure*}[t]
    \centering
    \includegraphics[width=0.33\textwidth]{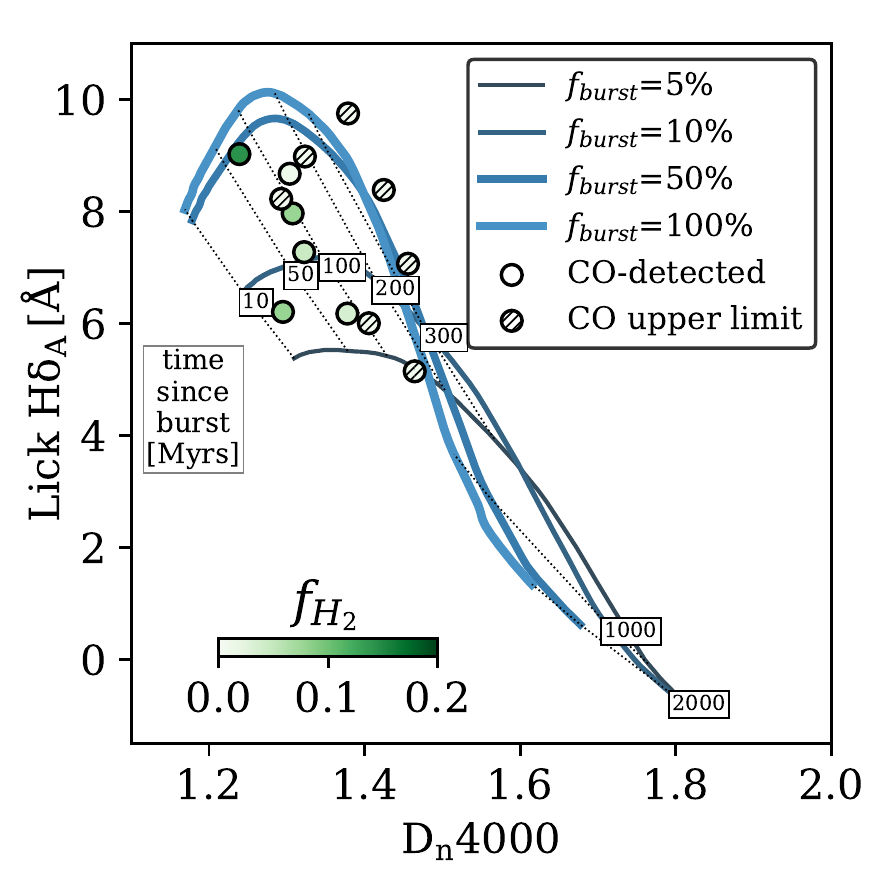}
    \includegraphics[width=0.66\textwidth]{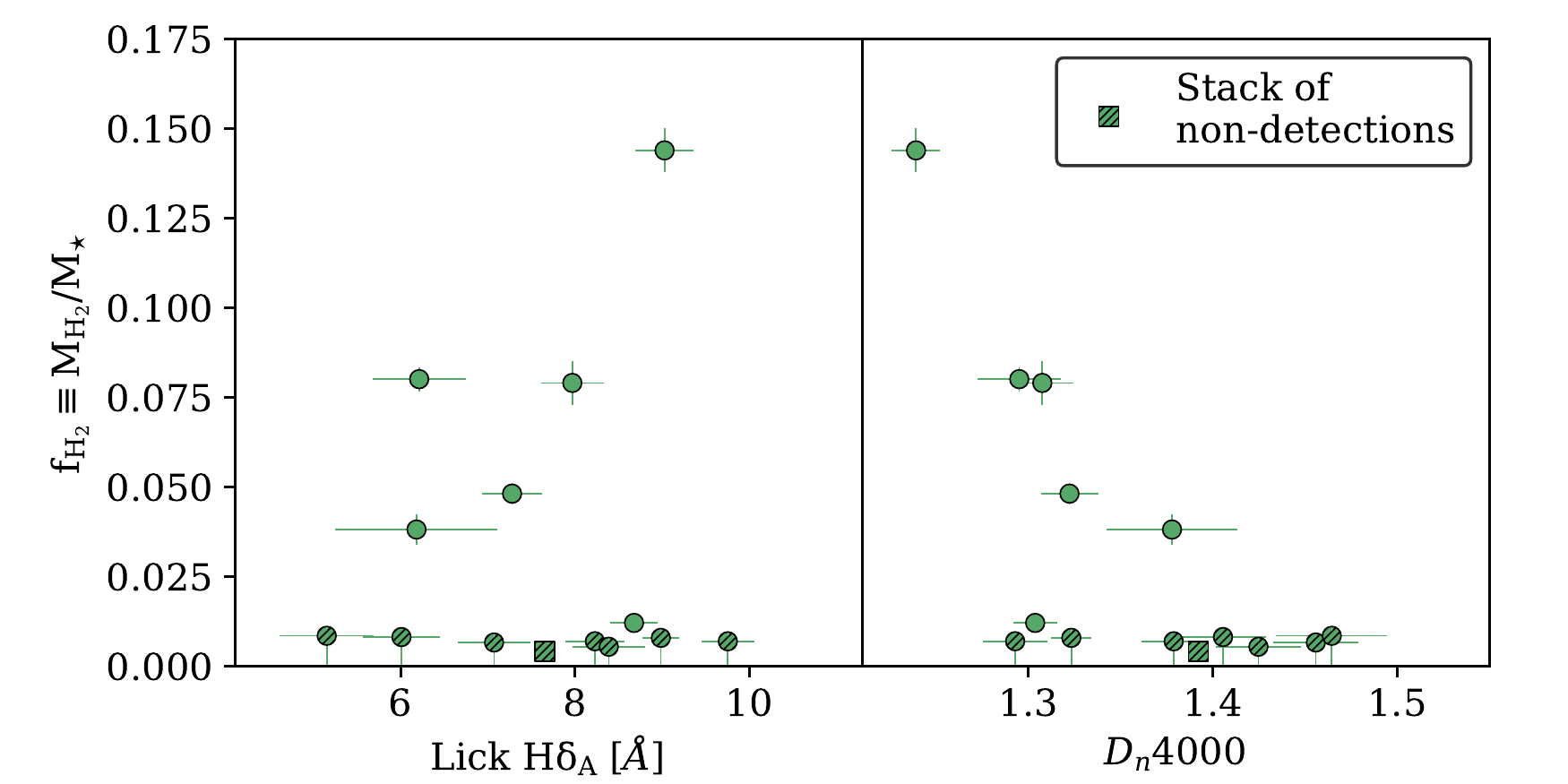}
    \caption{\squiggle galaxies in $H\delta_A$ versus $D_n4000$ space, colored by $H_2$ gas fraction, with upper limits indicated by triangles (left panel). Two-burst evolutionary tracks are indicated in blue, with lines of constant age in dotted black. Galaxies do not show a clear trend between $H_2$ reservoirs and $H\delta_A$ (center panel), however they exhibit higher $H_2$ gas fractions ($\gtrsim5\%$) at the lower $D_n4000$ (right panel). This suggests that molecular gas reservoirs diminish with time after $\sim100$ Myrs.}
    \label{fig:hd_d4000_samples}
\end{figure*}

It is also possible to define more inclusive scaling relations, along the lines of e.g., the fundamental metallicity relation \citep[e.g.,][]{mannucci:10}, that encompass the redshift-evolution of the galaxy population by spanning the range of properties through which galaxies evolve. In addition to being easier to graphically depict, the simplicity of such scaling relations is appealing because it does not require relying on evolving physics. One such parameter space is the relation between $H_2$ gas fraction (typically denoted as $f_{H_2}$ or $\mu$) and specific star formation rate, as shown in Figure \ref{fig:mu_ssfr}. In this figure we can include galaxies at a number of cosmic times, spanning from $z\sim0$ \citep{saintonge:11,saintonge:11b,saintonge:12} to $z\sim2$ \citep{tacconi:13}. At different redshifts, normal star-forming galaxies (all indicated by gray points) fall along a similar relation, but occupy varying regimes in sSFR. We show two such parameterizations of this integrated scaling from \citet{feldmann:20} and \citet{dou:21}. In this projection, we note that quiescent galaxies from \citet{spilker:18} and the stacked non-detections no longer fall outside of the scatter, rather they lie near lower-redshift star-forming counterparts. However, the $H_2$-rich galaxies in this sample remain dramatic outliers in this projection of gas fueling as well. We note that $z\sim0$ post-starburst galaxies from the \citet{french:18b} sample lie in a similarly offset location, but we omit them from the diagram for clarity of presentation.  This offset from all other samples of galaxies suggests that the deviation is temporary, but whether the H$_2$-rich galaxies rejuvenate at SFRs that correspond to their gas reservoirs or lose, heat, or deplete H$_2$ and quench permanently cannot be determined.

\subsection{Star Formation Histories and $H_2$ Reservoirs}

Finally, we investigate the relationship between the stellar populations of the galaxies in this sample and the residual gas fractions in Figure \ref{fig:hd_d4000_samples}. In Figure \ref{fig:hd_d4000_samples}a, we show \squiggle\ galaxies on the H$\delta_A$ versus $D_n4000$ space, colored by \fgas. Tracks were generated using { Flexible Stellar Population Synthesis} {\tt (FSPS)} \citep{conroy:09,conroy:10}, assuming two top hat star formation histories, $A_v=0.5$, and solar metallicity. Lines of constant age are indicated by thin dotted lines, and time since burst in Myr is labeled in rectangular boxes. Galaxies with the highest gas fractions (circular symbols) are located towards the left, or low $D_n4000$, portion of this panel, and galaxies that are undetected at the survey depths ($\lesssim3\times10^{9}M_{\odot}$, hatched circles) have higher $D_n4000$ values. This corresponds to roughly an inverse relationship between the existence of a gas reservoir and time since burst, with the majority of detectable $H_2$ gas in galaxies that quenched in the last ${\sim}100-200$ Myr. 

Trends between molecular gas fraction and stellar continuum indices are shown in the center and right panels of Figure \ref{fig:hd_d4000_samples}: $f_{H_2}$ versus H$\delta_A$ (Figure \ref{fig:hd_d4000_samples}b) and $D_n4000$ (Figure \ref{fig:hd_d4000_samples}c). There is no significant trend with H$\delta_A$ (Figure \ref{fig:hd_d4000_samples}b). This suggests that residual gas reservoirs are independent of the fraction of mass that was formed during the major burst, which primarily drives the maximum $H\delta_A$. The analysis of this trend is complicated by the fact that this parameter is double-valued in its time evolution. The trend with $D_n4000$ is much more apparent; the majority (5 of 7) of galaxies with $D_n4000<1.4$ have \fgas$>5\%$, while at higher values only one galaxy has detectable CO(2--1) emission. The average \fgas\ implied by the stacked CO(2--1) flux from individually undetected galaxies is consistent with the decreasing trend in \fgas\  with $D_n4000$. Unlike H$\delta_A$, $D_n4000$ increases monotonically with age; the trend we observe corresponds to declining H$_2$ reservoirs after the end of the starburst event. After about a hundred Myr the molecular gas fractions dwindle below $\sim1\%$. 
\begin{figure*}[!t]
    \centering
    \includegraphics[width=0.45\textwidth]{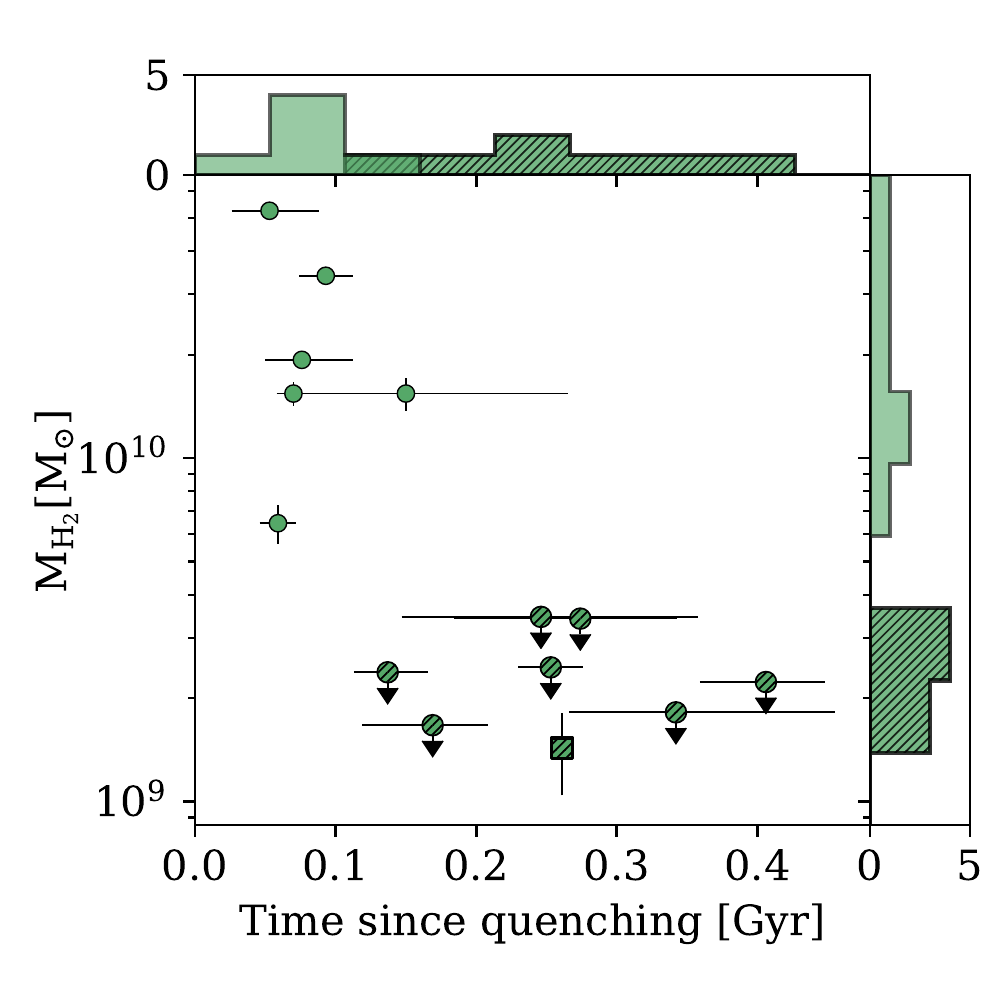}
    \includegraphics[width=0.45\textwidth]{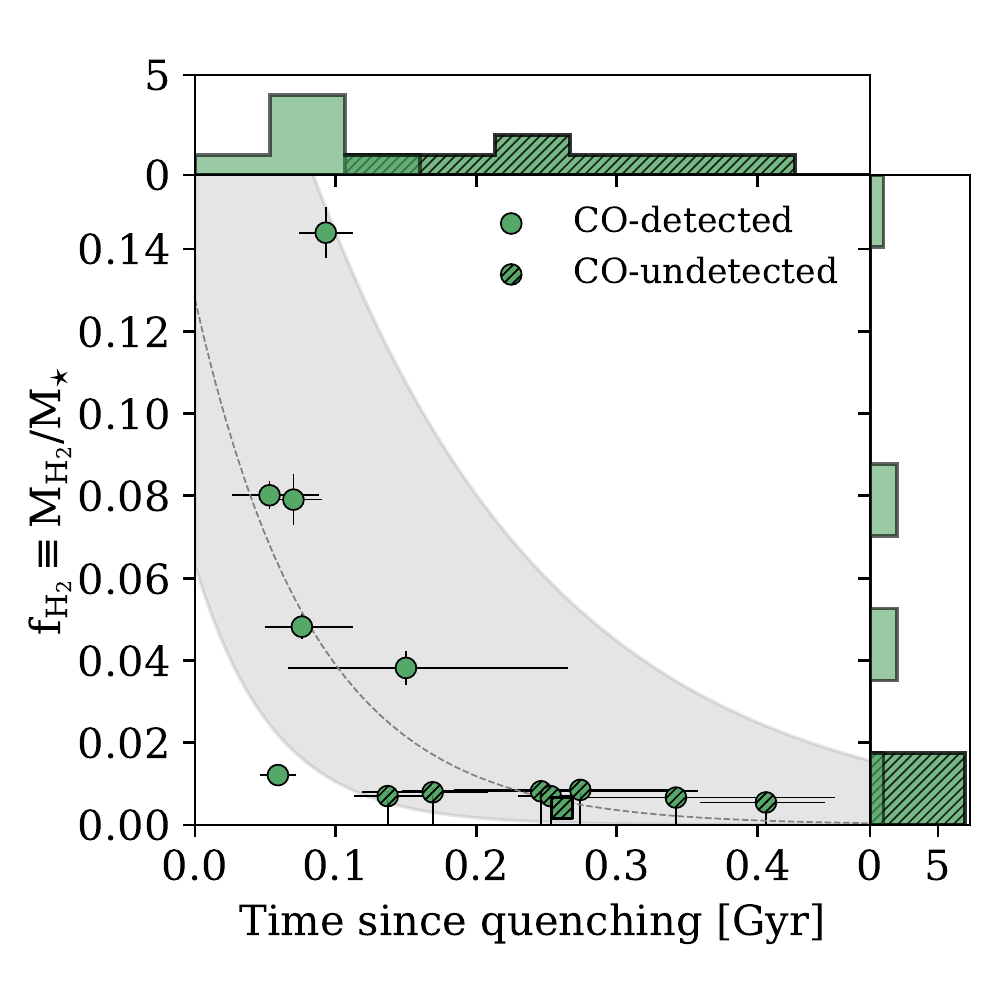}
    \caption{$H_2$ gas mass (left) and gas fraction (right) versus time since quenching for \squiggle targets (green circles) and average $H_2$ fraction derived from the stack of individually undetected galaxies (hatched square). { Histograms of the two samples are included at the top and right of each panel.} The correlation between whether a galaxy retains significant $H_2$ and the time since quenching is striking; all galaxies older than $t_q\gtrsim 200$ Myrs are undetected in CO emission. Best-fit exponentially declining relation (with $\tau_{dep}\approx84\pm45$ Myrs) is included as gray band in the right panel.}
    \label{fig:h2_tq}
\end{figure*}

The \texttt{Prospector} stellar population synthesis modeling used to determine e.g., the stellar masses and SFR of galaxies in the sample also provides flexible star formation histories. From these star formation histories, we compute the time between rapid quenching of star formation and the time of observation, which we refer to as the time since quenching ($t_q$; see Suess \etal 2021 subm.). In Figure \ref{fig:h2_tq}, we show that the $H_2$ gas masses (left) and fractions (right) are even more strongly correlated with $t_q$ than with the empirical indices. This is likely driven by the fact that these derived star formation histories are based on the full suite of spectral indices, essentially incorporating complex combinations of features that vary on different timescales (see e.g., Figure \ref{fig:hd_d4000_samples}a). It is immediately clear that the $H_2$-rich versus $H_2$-poor bimodality is statistically significant; only galaxies with $t_q\lesssim 200$ Myr retain $\fgas >1\%$.  { We note that the single galaxy that is detected in H$_2$, but has a slightly longer $t_q$ such that it overlaps with the H$_2$-poor subset is the single object that falls out of the DR14 spectroscopic S/N cuts. This may explain the relatively large uncertainty on $t_q$, however we include this potentially discrepant object in our analysis to avoid introducing a confirmation bias.} We perform a student T test { on the full sample} and verify that the detected $H_2$ and 3$\sigma$ limits on $M_{H_2}$ are not drawn from the same distribution, with p=0.02. In addition to the bimodality, we quantify the implied exponential decay timescale by fitting $\ln M_{H_2}/M_{\star}$ versus $t_q$, including the detected galaxies and the stack of undetected targets using Orthogonal distance regression with \texttt{scipy.ODR}. The resulting fit and confidence interval are included on \ref{fig:h2_tq}b as dashed line and gray band. This analysis yields an exponential depletion timescale of 84$\pm$45 Myr, a remarkably rapid decline in the H$_2$ reservoirs post-quenching, as we discuss further below.

\section{Discussion}\label{sec:discussion}

The primary result of this paper is that nearly half of the massive, post-starburst galaxies at $z\sim0.6$ in the \squiggle\ sample retain significant $H_2$ reservoirs (${\sim}1-5\times10^{10}M_{\odot}$), building upon our pilot study of two galaxies \citep{suess:17}. Timing derived from stellar population synthesis modeling suggests that this $H_2$ disappears rapidly; no CO(2--1) emission is detected in galaxies observed $\gtrsim$150 Myrs after their star formation truncated.  The existence of similar enigmatic molecular gas reservoirs has also been reported in a number of samples of post-starburst galaxies in the local Universe \citep[e.g.][]{french:15,rowlands:15,alatalo:16}. Moreover, we find a striking difference in the molecular gas properties of young vs. old post-starbursts, suggesting that the cold gas rapidly disappears $\sim$100-200 Myr post-quenching. Intriguingly, a detailed study of the star formation histories of the local sample found a similar anti-correlation between $t_q$ and the $H_2$ reservoirs, finding an exponential depletion timescale of 117-230 Myrs \citep{french:18b}. Those local samples span a much larger range in stellar masses than \squiggle\ ($9\lesssim \log \mstar/\msun {\lesssim} 11.5$). At low-redshift, similarly massive ($\log \mstar/\msun \sim$11) z$\sim$0 galaxies tend to be products of less extreme ($\sim$10\% mass fractions) bursts. Regardless, the similarity with the maximum $t_q$ beyond which no \squiggle galaxies are detected in CO(2--1) is striking, suggesting the two populations experience similar ties between quenching and cold gas depletion in the $\sim$200 Myrs after quenching.

\begin{figure}[!t]
    \centering
    \includegraphics[width=0.45\textwidth]{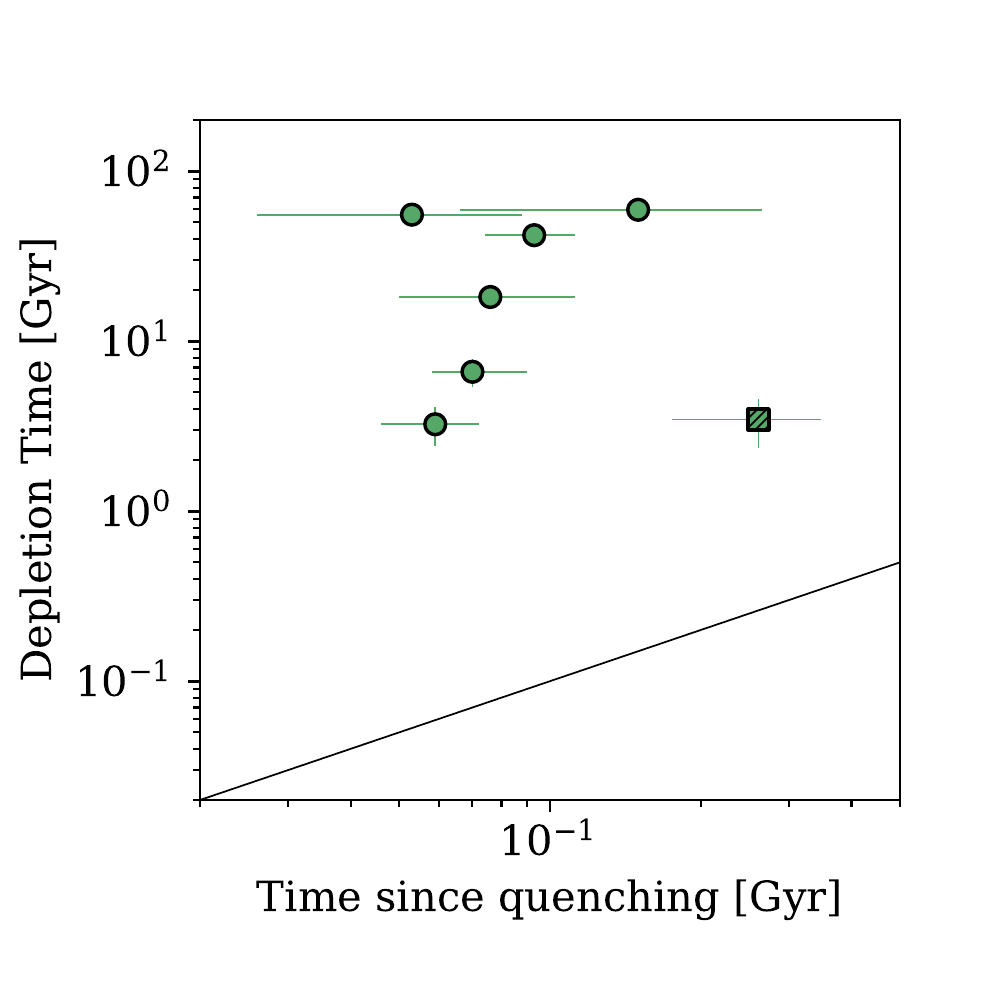}
    \caption{Depletion time versus time since quenching for CO-detected \squiggle targets (circles) and average depletion time derived from the stack of individually undetected galaxies (hatched square). Depletion times are much longer than the time since quenching (one-to-one line is indicated by the black line), necessitating additional physical heating or removal of $H_2$ beyond residual star formation to connect the younger and older groups if this represents and evolutionary sequence. }
    \label{fig:tdep}
\end{figure}

We emphasize that the depletion times due to ongoing star formation for the \squiggle galaxies are very long ($\gtrsim1$ Gyr) given their large $H_2$ reservoirs and low unobscured star formation rates. In Figure \ref{fig:tdep} we show the star formation depletion times ($\tdep\equiv\Mgas/SFR$) for \squiggle targets (green symbols) and stack of non-detections (square) versus $t_q$. It is immediately clear that the younger \squiggle galaxies have insufficient residual star formation for the observed reservoirs to deplete, especially if we assume that this trend corresponds to an evolutionary sequence.  The exponential depletion time calculated for this sample ($\tau_{dep}\sim$84$\pm$45 Myrs) is comparable (within $\sim$1$\sigma$) to the depletion timescale found by \citet{french:18b} in low-redshift post-starburst galaxies. \citet{french:18b} argue convincingly that such a rapid depletion timescale cannot be explained by on-going minimal star formation, extreme variations in stellar populations (e.g., the initial mass function), stellar winds or supernova feedback, pointing to this as possible smoking gun evidence for AGN feedback. We emphasize that we cannot rule out the possibility that star formation is only temporarily halted in the $H_2$-rich \squiggle galaxies, which are caught in the opposite transition, immediately prior to rejuvenation. In that case, the two halves of the \squiggle sample (gas-rich and gas-poor) may represent two different populations altogether depending on whether or not they will resume star formation in the future.

Another possible explanation for the large $H_2$ reservoirs that do not appear to fuel star formation could be that these galaxies harbor heavily dust-obscured star formation, causing depletion time estimates to be severely underestimated. We do not see evidence of continuum emission due to dust-obscured star formation in these galaxies in the 2mm ALMA data presented in this work\footnote{As noted in \S 2.1, one galaxy, SDSS\_J0753+2403 has detected non-thermal synchrotron emission, which we attribute to AGN activity}, placing a limit of SFR $\lesssim$50 $\mathrm{M_{\odot}\,yr^{-1}}$. Although a number of galaxies in the \squiggle dataset are detected by the Very Large
Array (VLA) Faint Images of the Sky at Twenty-Centimeters
(FIRST) survey \citep{first}, those data are too shallow to be sensitive to realistic levels of radio emission due to on-going star formation (such fluxes would correspond to $SFR\gtrsim1000 \mathrm{\msun\ yr^{-1}}$). Instead we interpret those detections as originating from AGN activity and explore that connection in a separate paper \citep{greene:20}. In general, estimating the SFR for post-starburst galaxies is challenging. An analysis of local E+A galaxies found significant scatter amongst different indicators, concluding that total infrared luminosity provides overestimates of the intrinsic SFRs \citep{smercina:18}. This implies that if SFR$_{IR} \lesssim 50\mathrm{M_{\odot}\,yr^{-1}}$ under standard assumptions, the true upper limit would be even stronger. We note that even if the SFRs are underestimated by an order of magnitude, the depletion times for the most extreme, youngest \squiggle would still be closer to $\sim1$Gyr, which is much longer than the range in $t_{q}$ probed by this sample. Therefore, if we assume that the trend in Figure \ref{fig:h2_tq} is a time sequence, $\sim$ 2 dex of dust-obscured star formation would be needed to deplete the $H_2$, which we would expect to see in continuum emission in the Band 4 ALMA data.

Interestingly, this empirical finding of large $H_2$ reservoirs in the youngest \squiggle galaxies is suggestively similar to an observed trend in the AGN occurrence rates within the same sample \citep{greene:20}. This is especially clear for AGN identified via optical emission lines (high \oiii$/\hbeta$ ratios), which are $\sim10$ times more common in the youngest \squiggle galaxies. Although the parent sample is the same for both studies, only one galaxy (SDSS\_J1448+1010) with strong \oiii{ \,emission,} indicating the presence of an AGN, is included in this ALMA sample. Therefore, we cannot make any robust claims about a possible causal correlation between the presence of an actively accreting supermassive black hole and the removal, heating, or destruction of $H_2$ within or surrounding these massive post-starburst galaxies.

While post-starburst galaxies remain a subdominant population of galaxies for the past $\sim7$ Gyrs, observational studies of massive galaxies of the high-redshift universe have begun to demonstrate that around $z\sim2-3$ the general population of massive ($\log M_*/M_{\odot}\sim 11$) galaxies is in the process of rapidly quenching their primary episode of star formation \citep{whitaker:12a, muzzin:13,tomczak:14,davidzon:17}.  Spectroscopic samples of such distant, massive galaxies indicate that post-starburst stellar populations are common \citep[e.g.][]{sande:11,sande:13,bezanson:13a,belli:15,carnall:19,kriek:19,wild:20, tacchella:21}, suggesting that many massive galaxies undergo a dramatic truncation of star formation. Which physical processes are responsible for driving those quenching events that can also concurrently destroy, deplete, or heat their molecular gas reservoirs is poorly constrained empirically. \citet{bezanson:19} placed stringent upper limits $f_{H_2}\lesssim 7\%$ in a $z=1.522$ galaxy and similarly \citet{williams:21} expanded the sample to include five additional galaxies. From the overall low $H_2$ fractions in that sample, \citet{williams:21} conclude the need for rapid ($\tau_{dep}\sim0.3$ Gyr) depletion. We note that this timescale cannot be compared directly to the $t_q$ parameter used in this study, which starts specifically \emph{after} star formation shuts down. The galaxies in those samples were significantly older, with a post-burst age closer to $\gtrsim1$ Gyr than the \squiggle galaxies that are detected in CO(2--1). Therefore, although the galaxies at that epoch are generally more gas-rich than local galaxies \citep[e.g.,][]{tacconi:13, tacconi:18, freundlich:19, belli:21}, we would not have predicted that they would be detected based on the results presented in this paper. Therefore, while the striking similarity between studies of post-starburst galaxies at $z\sim0$ and the more extreme cases at $z\sim 0.6$ presented in this paper is suggestive that quenching does not necessarily coincide with the elimination of the $H_2$ reservoirs, direct studies of galaxies during the primary quenching epoch will be critical in strengthening our understanding of the quenching mechanisms at play.

While no model includes the existence of $H_2$ after star formation shuts down by construction, the rapid disappearance of these cold gas reservoirs in post-starburst galaxies is easily consistent with the gas-poor nature of older quiescent galaxies \citep[e.g.,][]{young:11,young:14}. As they stand, these data present a challenge to galaxy formation models. Larger, more statistical studies of $H_2$ reservoirs in this and similar populations of recently quenched galaxies could precisely time the disappearance of $H_2$. We anticipate that spatially resolved $H_2$ maps may provide clues as to the distribution and kinematics of the $H_2$, which may help explain how it is stabilized against collapse. Initial studies of the stellar kinematics of this sample have revealed a range of rotational support \citep{setton:20} and, in one case, that the CO(2--1) kinematics follow the stellar motion \citep{hunt:18}. \citet{french:18b} found a smooth exponential depletion of gas over time to provide the best fit for local galaxies. Deeper observations of individually undetected galaxies could assess how smooth the transition from $H_2$-rich is for our intermediate-redshift galaxies, as the current observations seem to reveal a more discontinuous distribution. 

{ Theoretical and observational studies have pointed out that dynamical support against H$_2$ collapse could also stem from turbulent pressure in the gas. For example, slightly lower star formation efficiency is observed in morphologically classified local early type galaxies; the same molecular gas reservoirs form $\sim$2.5 times fewer stars in ellipticals than spiral galaxies \citep[e.g.,][]{davis:14}. Similarly, quiescence below the main sequence is driven by both depleted gas reservoirs and diminished star formation efficiency \citep[e.g.,][]{piotrowska:20}. This effect is often attributed to relatively deep gravitational wells and increased disordered orbits, ``morphological quenching'' \citep{martig:09}, but could also be driven by streaming motions \citep{meidt:13}. \citet{gensior:20} found that the existence of compact, spheroidal structures can indeed drive turbulent pressure, pointing towards the feasibility of morphological quenching. This could be especially relevant given that the sizes of post-starburst galaxies are often compact, even with respect to older quiescent galaxies \citep[e.g.,][D. Setton, in prep.]{yano:16, wu:20}. It has been demonstrated the turbulent motions in the interstellar medium, perhaps induced by a combination of shocks and magnetic fields, could support H$_2$ against collapse \citep[e.g.,][]{federrath:15}. Simulations of relativistic jets can drive shocks that diminish star formation rates by a factor of $\sim2$ \citep[e.g.][]{mandal:21}, but not all specific implementations of jet astrophysics can effectively quench star formation \citep[e.g.,][]{su:21}. Furthermore, given that most of these processes can decrease star formation efficiency by only a modest factor of a few, it remains unclear whether even combined these models could explain the order of magnitude offsets in star formation efficiency exhibited by the youngest \squiggle\ galaxies. Perhaps further analysis of systematic suites of simulations \citep[e.g.,][]{su:21} could use the immediate decrease in star formation efficiency and subsequent rapid ($\sim100$ Myr) disappearance of H$_2$ to differentiate amongst feedback models. Such studies could be especially constrained by extracting observed quantities matched to the \squiggle\ dataset. In particular, \citet{su:21} showed the promising efficiency of cosmic ray jets, which rapidly shut off star formation on a similar timescale; a more careful comparison would be needed to assess whether the depletion time lag is also consistent.}

Although the similarity of the \citet{french:18b} results at $z\sim0$ and the more extreme post-starbursts at intermediate redshifts presented in this paper suggests a fundamental challenge to galaxy formation models, the strongest test will come from earlier times, at the peak epoch of galaxy quenching and transformation \citep[e.g.][]{wild:16}. In this paper, we focus on CO-based measurements of $H_2$, however recent studies have suggested elevated average cold gas reservoirs in quiescent galaxies at cosmic noon based on stacked far-infrared dust continuum emission \citep[e.g.][]{gobat:18,magdis:21}, in apparent contradiction with low or absent $M_{H_2}$ in individual galaxies \citep[e.g.][]{caliendo:21,williams:21,whitaker:21}. One possible interpretation is that these stacks include a subset of $H_2$-rich recently quenched ($t_q\lesssim200$ Myr) galaxies averaged with a depleted majority. Identifying these young galaxies requires spectroscopic data of sufficient quality to precisely measure star formation histories. Although current spectroscopic samples of galaxies at cosmic noon are somewhat rare, in the coming years massively multiplexed spectrographs with NIR capabilities, like the Prime Focus Spectrograph on Subaru \citep{takada:14} or MOONS {(Multi-Object Optical and Near-infrared Spectrograph)}) \citep{maiolino:21}, will produce hundreds of recently quenched targets at this critical epoch. Follow-up studies of their $H_2$ reservoirs, either with CO or dust-based estimates, will ultimately determine the timescale and simultaneity of quenching and the disappearance of $H_2$ in massive quiescent galaxies.

\begin{acknowledgments}
RSB, JEG, DJS, and DN gratefully acknowledge support from NSF-AAG\#1907697, \#1907723 and \#1908137. JSS acknowledges support provided by NASA through NASA Hubble fellowship grant \#HF2-51446 awarded by the Space Telescope Science Institute, which is operated by the Association of Universities for Research in Astronomy, Inc., for NASA, under contract NAS5-26555. RF acknowledges financial support from the Swiss National Science Foundation (grant no PP00P2\_157591, PP00P2\_194814, and 200021\_188552). This paper makes use of the following ALMA data: ADS/JAO.ALMA \#2016.1.01126.S and ADS/JAO.ALMA \#2017.1.01109.S. ALMA is a partnership of ESO (representing its member states), NSF (USA) and NINS (Japan), together with NRC (Canada), MOST and ASIAA (Taiwan), and KASI (Republic of Korea), in cooperation with the Republic of Chile. The Joint ALMA Observatory is operated by ESO, AUI/NRAO and NAOJ. The National Radio Astronomy Observatory is a facility of the National Science Foundation operated under cooperative agreement by Associated Universities, Inc. KAS gratefully acknowledges the UCSC Chancellor's Fellowship. 

This work was performed in part at the Aspen Center for Physics, which is supported by National Science Foundation grant PHY-1607611. We also thank the North American ALMA Science Center (NAASC) for their generous funding that helped support the participation of junior scientists (including DJS and KAS) at the Aspen Center for Physics workshop in February 2020. The NAASC is part of the National Radio Astronomy Observatory, a facility of the National Science Foundation operated under cooperative agreement by Associated Universities, Inc.

Funding for SDSS-III has been provided by the Alfred P. Sloan Foundation, the Participating Institutions, the National Science Foundation, and the U.S. Department of Energy Office of Science. The SDSS-III web site is http://www.sdss3.org/.

SDSS-III is managed by the Astrophysical Research Consortium for the Participating Institutions of the SDSS-III Collaboration including the University of Arizona, the Brazilian Participation Group, Brookhaven National Laboratory, Carnegie Mellon University, University of Florida, the French Participation Group, the German Participation Group, Harvard University, the Instituto de Astrofisica de Canarias, the Michigan State/Notre Dame/JINA Participation Group, Johns Hopkins University, Lawrence Berkeley National Laboratory, Max Planck Institute for Astrophysics, Max Planck Institute for Extraterrestrial Physics, New Mexico State University, New York University, Ohio State University, Pennsylvania State University, University of Portsmouth, Princeton University, the Spanish Participation Group, University of Tokyo, University of Utah, Vanderbilt University, University of Virginia, University of Washington, and Yale University.

{ The Legacy Surveys consist of three individual and complementary projects: the Dark Energy Camera Legacy Survey (DECaLS; Proposal ID \#2014B-0404; PIs: David Schlegel and Arjun Dey), the Beijing-Arizona Sky Survey (BASS; NOAO Prop. ID \#2015A-0801; PIs: Zhou Xu and Xiaohui Fan), and the Mayall z-band Legacy Survey (MzLS; Prop. ID \#2016A-0453; PI: Arjun Dey). DECaLS, BASS and MzLS together include data obtained, respectively, at the Blanco telescope, Cerro Tololo Inter-American Observatory, NSF's NOIRLab; the Bok telescope, Steward Observatory, University of Arizona; and the Mayall telescope, Kitt Peak National Observatory, NOIRLab. The Legacy Surveys project is honored to be permitted to conduct astronomical research on Iolkam Du'ag (Kitt Peak), a mountain with particular significance to the Tohono O'odham Nation.

NOIRLab is operated by the Association of Universities for Research in Astronomy (AURA) under a cooperative agreement with the National Science Foundation.

This project used data obtained with the Dark Energy Camera (DECam), which was constructed by the Dark Energy Survey (DES) collaboration. Funding for the DES Projects has been provided by the U.S. Department of Energy, the U.S. National Science Foundation, the Ministry of Science and Education of Spain, the Science and Technology Facilities Council of the United Kingdom, the Higher Education Funding Council for England, the National Center for Supercomputing Applications at the University of Illinois at Urbana-Champaign, the Kavli Institute of Cosmological Physics at the University of Chicago, Center for Cosmology and Astro-Particle Physics at the Ohio State University, the Mitchell Institute for Fundamental Physics and Astronomy at Texas A\&M University, Financiadora de Estudos e Projetos, Fundacao Carlos Chagas Filho de Amparo, Financiadora de Estudos e Projetos, Fundacao Carlos Chagas Filho de Amparo a Pesquisa do Estado do Rio de Janeiro, Conselho Nacional de Desenvolvimento Cientifico e Tecnologico and the Ministerio da Ciencia, Tecnologia e Inovacao, the Deutsche Forschungsgemeinschaft and the Collaborating Institutions in the Dark Energy Survey. The Collaborating Institutions are Argonne National Laboratory, the University of California at Santa Cruz, the University of Cambridge, Centro de Investigaciones Energeticas, Medioambientales y Tecnologicas-Madrid, the University of Chicago, University College London, the DES-Brazil Consortium, the University of Edinburgh, the Eidgenossische Technische Hochschule (ETH) Zurich, Fermi National Accelerator Laboratory, the University of Illinois at Urbana-Champaign, the Institut de Ciencies de l'Espai (IEEC/CSIC), the Institut de Fisica d'Altes Energies, Lawrence Berkeley National Laboratory, the Ludwig Maximilians Universitat Munchen and the associated Excellence Cluster Universe, the University of Michigan, NSF's NOIRLab, the University of Nottingham, the Ohio State University, the University of Pennsylvania, the University of Portsmouth, SLAC National Accelerator Laboratory, Stanford University, the University of Sussex, and Texas A\&M University.

BASS is a key project of the Telescope Access Program (TAP), which has been funded by the National Astronomical Observatories of China, the Chinese Academy of Sciences (the Strategic Priority Research Program ``The Emergence of Cosmological Structures'' Grant \# XDB09000000), and the Special Fund for Astronomy from the Ministry of Finance. The BASS is also supported by the External Cooperation Program of Chinese Academy of Sciences (Grant \# 114A11KYSB20160057), and Chinese National Natural Science Foundation (Grant \# 11433005).

The Legacy Survey team makes use of data products from the Near-Earth Object Wide-field Infrared Survey Explorer (NEOWISE), which is a project of the Jet Propulsion Laboratory/California Institute of Technology. NEOWISE is funded by the National Aeronautics and Space Administration.

The Legacy Surveys imaging of the DESI footprint is supported by the Director, Office of Science, Office of High Energy Physics of the U.S. Department of Energy under Contract No. DE-AC02-05CH1123, by the National Energy Research Scientific Computing Center, a DOE Office of Science User Facility under the same contract; and by the U.S. National Science Foundation, Division of Astronomical Sciences under Contract No. AST-0950945 to NOAO.}

\end{acknowledgments}

\vspace{5mm}
\facilities{ALMA}

\software{astropy \citep{astropy}, scipy \citep{scipy}, seaborne \citep{seaborn}}

\end{document}